\documentclass[journal]{IEEEtran}
\usepackage{amsmath,amsfonts}
\usepackage{algorithmic}
\usepackage{algorithm}
\usepackage{array}
\usepackage{url}
\usepackage[colorlinks]{hyperref}
\hypersetup{citecolor=blue}
\usepackage{tabularx}
\usepackage{booktabs}
\usepackage{graphicx}
\usepackage{amssymb}
\usepackage{relsize}
\begin{document}
\title{\LARGE Beyond Covariance: Generative Spatial Correlation Modeling \\and Channel Interpolation for Fluid Antenna Systems}
\author{Zhentian Zhang, 
            Hao Jiang,~\IEEEmembership{Senior Member,~IEEE}, 
            Kai-Kit Wong,~\IEEEmembership{Fellow,~IEEE},\\
            Hyundong Shin, \emph{Fellow, IEEE}, and 
            Ross Murch, \emph{Fellow, IEEE}
\vspace{-10mm}

\thanks{The work of K. K. Wong is partly supported by the Engineering and Physical Sciences Research Council (EPSRC) under grant EP/W026813/1.}
\thanks{The work of H. Jiang is partly supported in part by the National Natural Science Foundation of China (NSFC) projects (No. 62471238).} 
\thanks{The work of H. Shin is supported by the National Research Foundation of Korea (NRF) grant funded by the Korean government (MSIT) (RS-2025-00556064 and RS-2025-25442355), and by the Ministry of Science and ICT (MSIT), Korea, under the ITRC (Information Technology Research Center) support program (IITP-2025-RS-2021-II212046), supervised by the IITP (Institute for Information \& Communications Technology Planning \& Evaluation).}
\thanks{The work of R. Murch was supported by the Hong Kong Research Grants Council Area of Excellence Grant AoE/E-601/22-R.}

\thanks{Z. Zhang and H. Jiang are with the National Mobile Communications Research Laboratory, Southeast University, Nanjing, 210096, China and H.~Jiang is also with the School of Artificial Intelligence, Nanjing University of Information Science and Technology, Nanjing 210044, China (e-mail: zhentianzhangzzt@gmail.com, jianghao@nuist.edu.cn).}
\thanks{K. K. Wong is affiliated with the Department of Electronic and Electrical Engineering, University College London, Torrington Place, WC1E 7JE, United Kingdom and he is also affiliated with the Department of Electronic Engineering, Kyung Hee University, Yongin-si, Gyeonggi-do 17104, Korea.}
\thanks{H. Shin is with the Department of Electronics and Information Convergence Engineering, Kyung Hee University, Yongin-si, Gyeonggi-do 17104, Republic of Korea (e-mail: hshin@khu.ac.kr).}
\thanks{R. Murch is with the Department of Electronic and Computer Engineering, Hong Kong University of Science and Technology, Clear Water Bay, Hong Kong SAR, China (e-mail: eermurch@ust.hk).}

\thanks{\em Corresponding authors: H. Jiang}
}

%

\maketitle

\begin{abstract}
Fluid antenna systems (FAS) enable unprecedented spatial diversity within a compact form factor by flexibly switching among high-density antenna ports. To activate this capability, channel state information (CSI) over the ports is required, which implies high estimation overhead because the number of ports is usually very large. Conventional estimation schemes tend to first estimate the CSI for a small number of ports and then infer the CSI for the remaining antenna ports by interpolation exploiting correlation characteristics. However, existing correlation-based techniques lack generalization ability, and the fundamental limits of interpolating the CSI from sparse observations remain poorly understood. This paper adopts a generative modeling framework for characterizing the channel correlation among the FAS ports that departs fundamentally from covariance-descriptive models. Specifically, we represent the spatially sampled channel as a $p$th-order autoregressive (AR) Gauss-Markov process, which provides a principled and tunable tradeoff between model complexity and approximation accuracy via the AR order. In so doing, we can characterize the limits of channel interpolation by deriving the globally optimal minimum mean-square error (MMSE) estimator and establishing a tight lower bound on the minimum number of observations required to meet a prescribed reconstruction error. To reduce the complexity of MMSE estimation, we then exploit the state-space structure due to the ${\rm AR}(p)$ model and develop a Kalman filtering/smoothing-based interpolation algorithm. The resulting method attains the optimal MMSE performance with strictly linear complexity $\mathcal{O}(N)$ with $N$ denoting the number of ports, resulting in a scalable, efficient, and theoretically grounded framework for practical FAS channel reconstruction.
\end{abstract}

\begin{IEEEkeywords}
Fluid antenna system (FAS), channel correlation, channel reconstruction, fundamental limits.
\end{IEEEkeywords}

\vspace{-3mm}
\section{Introduction}
\subsection{Background}
\IEEEPARstart{A}{mong all diversity} techniques, spatial diversity is uniquely attractive because it comes without bandwidth expansion and does not require scheduling \cite{MIMO1,Telatar-1999}. As long as space is available, it can be utilized to enhance reliability and create capacity. In conventional multiple-input multiple-output (MIMO) systems, spatial diversity is obtained by deploying multiple fixed-position antennas (FPAs), far apart from each other, at the transmitter and receiver sides. The idea is that by having antennas at distant locations, their received signals are uncorrelated and the probability that all the antennas receive poor signals will be much lower. Multiuser MIMO even turns wireless channels into orthogonal data pipes to accommodate multiple users in the spatial domain \cite{wong2000opt,wong2002per} and become the backbone of the physical layer in the fifth generation (5G) \cite{MIMO2,Villalonga2022spectral}. Many have speculated that the sixth generation (6G) will see many more antennas at a base station (BS) in order to satisfy the growing demands \cite{10379539}. However, increasing antenna count is extremely difficult. It not only increases the overheads for channel estimation and precoding optimization but also the power consumption of the BS due to the increasing peak-to-average power ratio that significantly reduces the efficiency of power amplifiers \cite{Hung-2014}. Moreover, further increasing the size of MIMO increases the cost of radio-frequency (RF) chains.

MIMO systems are groundbreaking but the time has come to go beyond. In this context, {\em reconfigurable antennas} can be the game changer \cite{Bernhard-2007} and in \cite{Wong-2020cl,FAS}, Wong {\em et al.}~introduced the fluid antenna system (FAS) that utilizes antenna position reconfigurability on a given aperture for spatial diversity and revealed that extraordinary diversity can be obtained even in a compact space. More precisely, FAS is actually a hardware-agnostic system concept that treats the antenna as a reconfigurable physical-layer resource to broaden system design \cite{FAS_Survey,FAS_survey_Hong,Lu-2025,FAS_enabler,FAS_wu_tuo1}. In practice, FAS hardware may come in the form of movable elements \cite{Zhu-Wong-2024}, liquid antennas \cite{shen2024design,Shamim-2025}, reconfigurable pixels \cite{zhang2024pixel,tong-2025pixel}, metasurfaces \cite{Zhang-jsac2026,Liu-2025arxiv}, etc.

In recent years, the interest of FAS has surged and much progress has been made to study the diversity benefits \cite{Jakes1,Jakes2}, simplify performance analysis through channel modeling \cite{Clarke3,Clarke4}, develop new multiple access approaches \cite{FAS_MA1,FAS_MA2}, improve signal processing \cite{FAS_AD,FAS_wu_tuo2}, apply it for sensing \cite{FAS_ISAC,FAS_Array_Design}, to name just a few. There are many applications that can benefit from the use of FAS, see \cite[Section VIII]{FAS_Survey}.

What is common in the prior works is the need of channel state information (CSI) for FAS to function effectively. CSI is essential so that FAS can utilize the most advantageous port(s) for communication. CSI estimation in FAS has been studied recently in \cite{CSI2,Aven_oversampling,CSI0,CSI3,CSI1}. Due to the large number of ports in FAS, channel estimation normally involves:
\begin{enumerate}
\item {\bf CSI sampling}---Uniformly sample a number of antenna ports over the FAS aperture and estimate their CSI.
\item {\bf CSI reconstruction}---Estimate the CSI of the remaining ports using interpolation or reconstruction. 
\end{enumerate}
But fundamentally, little is understood about this process.

\vspace{-3mm}
\subsection{Related Works}
{\em Sampling and Correlation Modeling}---A crucial observation of FAS is that the channel in the space continuum is inherently correlated. Different from the standard half-wavelength placement rule in FPAs, FAS promotes oversampling within a given aperture, which requires rigorous correlation modeling to accurately represent the FAS channels. Classical correlation modelings such as Jakes' \cite{Jakes0} and Clarke's \cite{Clarke0,Clarke2} model are widely adopted in the case of rich isotropic scattering. In spite of this, performance analysis under these models is {\em prohibitively challenging} due to the high-dimensional structures. 

{\em Correlation Approximation}---To regain analytical tractability, one could simplify the channel correlation structures via approximation. It was reported in \cite[Fig.~6]{FAS_Survey}, \cite[Fig.~3]{Clarke3} that sparsity exists within the span of eigenvalues where there are only few dominant eigenvalues in the correlation matrix of an extremely dense array with finite aperture. This has led to the spatial block-correlation model \cite{Clarke3} which greatly simplifies the derivation of cumulative density function (CDF), allowing tradeoff between accuracy and complexity. Alternatively, \cite{Aven_oversampling} considered an electromagnetic-compliant channel model which deviates from classical correlation modeling.

{\em Channel Reconstruction}---To perform reconstruction, current approaches \cite{CSI0,CSI1} tend to first {\em estimate} a partial subset of available ports and then {\em interpolate} the whole channels via different correlation factors. For instance, \cite{CSI2} interpolated the estimated observations using the multipath channel modeling with finite scatterers in which all ports share a deterministic structure in angular domain. Differently, \cite{CSI3} directly learned the correlation between ports via data-driven machine-learning techniques and interpolated the whole channel simultaneously. Despite the progress, the fundamental reconstruction limits in relation to the correlation structures are not understood.

\vspace{-3mm}
\subsection{Challenges}
{\em Lack of Flexibility in Correlation Approximation}---Though dimension collapse through sparsity within eigenvalue space such as the block-correlation approximation in \cite{Clarke3} is useful, this line-of-work relies heavily on sorting out effective eigenvalues by well-tuned thresholding. On the other hand, electromagnetic compliant modeling \cite{Aven_oversampling} requires strictly established physical radiation structure before offering any insights, which would pose another challenges.

{\em Interpolation Limits}---Under the approach of partial observation estimation, there are fundamental questions:
\begin{itemize}
\item {\em ``How port selection strategies will affect interpolation?''}
\item {\em ``How many observations are adequate?''}
\end{itemize}
In \cite{Aven_oversampling}, it was shown that {\em spatial oversampling is essential}. But whether even sampling is optimal, and the fundamental  limits for CSI reconstruction (i.e., interpolation accuracy for a given number of observations), are not known.

{\em Design Principle}---Complex radiating environment makes structural channel modeling less appealing. Adaptive interpolation via a trainable process can provide a robust solution but inconsistency between exact model and randomized observations brings challenges to practical algorithmic designs. 

\vspace{-3mm}
\subsection{Contributions}
This paper has made the following contributions:
\begin{itemize}
\item {\em Correlation Approximation by Generative Modeling}---Different from conventional correlation descriptive modeling, we adopt a {\em generative modeling} approach for correlation approximation through a $p$-order auto-regression (AR) series. In this way, one can balance the approximation complexity and accuracy by adjusting the AR($p$)'s order. Additionally, we verify that the AR($p$)-modeling is generally equivalent to a Gauss-Markov process. Moreover, the proposed AR($p$)-based correlation representation also naturally accommodates the interpolation process, guiding practical interpolation algorithm designs.
\item {\em Fundamental Limits for Interpolation}---In this work, the relationship between channel interpolation and correlation modeling is clarified mathematically. We illustrate that with perfect correlation modeling, full reconstruction based on {\em arbitrary observations} is possible. Therefore, the interpolation error floor will be limited by the accuracy of the correlation approximation. Via the joint Gaussian distribution (correlation model), the {\em globally optimal solution} to interpolation under a minimum mean square error (MMSE) estimator is provided. Moreover, we also clarify the fundamental lower-bound for the {\em minimum observation count} admitting a prescribed interpolation error target theoretically, which is proved to be extremely tight to the practical MMSE estimator.
\item {\em Linear-Complexity and Optimal Interpolation Designs}---The MMSE estimator requires computational complexity in the cubic order to the number of ports, which is highly impractical because FAS typically has many ports. Thus, efficient alternatives to the MMSE estimator is sought. Specifically, a Kalman filtering and smoothing estimator is introduced based on the AR($p$) modeling, yielding MMSE estimation under the linear state-space AR($p$) models but requiring only linear-complexity order.
\end{itemize}


The rest of this paper is structured as follows. In Section \ref{sec:ARp_approx}, correlation modeling via the AR($p$) Gauss-Markov process is elaborated, which is validated theoretically in Section \ref{subsec:ARmax_dist}. Then the fundamental principles between the correlation model and interpolation via arbitrary sparse observations within a fixed FAS aperture and practical algorithm designs are introduced in Section \ref{sec:channel_completion}. Numerical results are illustrated in Section \ref{sec.numerical} and finally, conclusions are drawn in Section \ref{sec.conclusion}. 

{\em Notations:} We use the following notations throughout the paper. Scalars are denoted by italic letters (e.g., $x$), vectors by bold lowercase letters (e.g., $\mathbf x$), and matrices by bold uppercase letters (e.g., $\mathbf X$). The operators $(\cdot)^\top$, $(\cdot)^\mathsf H$, $\mathrm{tr}(\cdot)$, $\|\cdot\|$, and $\|\cdot\|_\mathrm F$ denote transpose, Hermitian transpose, trace, Euclidean norm, and Frobenius norm, respectively; $\mathbf I$ denotes the identity matrix. Additionally, the notation $\mathcal{CN}(\boldsymbol\mu,\mathbf C)$ represents a circularly symmetric complex Gaussian distribution with mean $\boldsymbol\mu$ and covariance $\mathbf C$, and $\mathbb E[\cdot]$ is expectation.

\vspace{-2mm}
\section{AR($p$) Gauss-Markov Approximation}\label{sec:ARp_approx}
\subsection{Exact Correlation Modeling via Clarke's Model}
Our starting point is the Clarke's correlation model \cite{Clarke0}, which has been shown to be accurate under rich scattering \cite{Clarke2}. The model has also been adopted widely in the literature. Let $\boldsymbol{g} \in \mathbb{C}^{N}$ denote the channel vector with $N$ ports evenly placed over a linear space of $W\lambda$ with $\lambda$ being the carrier wavelength, and let $\boldsymbol{\Sigma} \in \mathbb{C}^{N \times N}$ be the corresponding spatial correlation matrix. Based on the even placement of ports and the assumption of Clarke's correlation model, the matrix $\boldsymbol{\Sigma}$ takes the form of a Toeplitz matrix \cite[(11)]{Clarke3}, \cite[(3)]{Clarke4}
\begin{equation}\label{eq.Clarke}
\boldsymbol{\Sigma}=\begin{pmatrix}
a(0)&a(1)&\dots&a(N-1)\\
a(-1)&a(0)&\dots&a(N-2)\\
\vdots&\vdots&\ddots\\
a(-N+1)&a(-N+2)&\dots&a(0)
\end{pmatrix},
\end{equation}
where the generation function is $ a(\ell)=\operatorname{sinc}\left(\frac{2\pi \ell W}{N-1}\right)$. 

With the correlation matrix $\boldsymbol{\Sigma}$, one could generate $\boldsymbol{g}$ feasibly via eigenvalue-based construction
\begin{equation}\label{eq:channel_eigenvalue}
\boldsymbol{g}=\boldsymbol{U}\boldsymbol{\Lambda}^{\frac{1}{2}}\boldsymbol{g}_{0},
\end{equation}
where $\boldsymbol{U}$ is the matrix of eigenvectors and $\boldsymbol{\Lambda}$ is the diagonal matrix of eigenvalues obtained from the eigenvalue decomposition of the correlation matrix $\boldsymbol{\Sigma} = \boldsymbol{U}\boldsymbol{\Lambda}\boldsymbol{U}^{\mathrm{H}}$, and $\boldsymbol{g}_{0} \in \mathbb{C}^{N}$ is randomly generated from $ \mathcal{CN}\left(\boldsymbol{0},\sigma^2{\bf I}\right)$.

Under the Clarke's Toeplitz model, the port-indexed channel vector $\boldsymbol{g} \sim \mathcal{CN}(\boldsymbol{0},\boldsymbol{\Sigma})$ follows a full $N$-dimensional complex Gaussian distribution whose spatial dependence is encoded by the correlation sequence $\{a(\ell)\}_{\forall\ell}$ in \eqref{eq.Clarke}. Although random channel realizations can be generated efficiently via \eqref{eq:channel_eigenvalue}, many FAS performance metrics (e.g., selection gain distributions, tail probabilities, etc.) depend upon {\em the joint $N$-dimensional distribution and are consequently analytically intractable.}

To regain tractability while preserving the essential correlation structure, correlation modeling approximation becomes crucial especially for cases in which the explicit probability density function (PDF) and CDF expressions are required. For instance, by exploiting the sparsity in the eigenvalue domain \cite[Fig.~6]{FAS_Survey}, block-correlation approximation reduces the variables' dimension by approximating the covariance matrix \cite[(20)]{Clarke3}, facilitating tractable analyses via simplified PDF and CDF. In \cite[(17)]{Clarke4}, the authors utilized moment matching to further simplify the expressions. Essentially, {\em dimension reduction} is the prevailing principle behind correlation modeling and approximation, where one seeks low-rank or compressed representations of the covariance structure.

 In this work, a fundamentally different approach is adopted. We approximate the port-indexed process $\{g_k\}_{k=1}^N$ by a finite-memory (order-$p$) complex Gauss-Markov model. Instead of approximating the correlation matrix as a static object in the statistical domain, we impose a dynamic generative structure on the process itself, whereby correlation emerges from a recursive state evolution. In other words, rather than compressing the covariance, we re-parameterize the correlation through a finite-order state-space mechanism. Note that while this paper focuses on the Clarke's model as an example, our proposed approach is applicable for any correlation structures. 

\vspace{-2mm}
\subsection{AR($p$) Gauss-Markov Correlation Modeling}\label{subsec:ARp_model}
To do so, we leverage the fact that the port-indexed channel process $\{g_k\}_{k=1}^N$ is {\em wide-sense stationary} and exhibits localized spatial correlation along the array. Specifically, the wide-sense stationarity indicates the spatial homogeneity of the uniform linear array, whereby second-order statistics depend only on port separation rather than absolute index.  For example, the covariance matrix in \eqref{eq.Clarke} offers a direct illustration of wide-sense stationarity in the spatial (port-indexed) domain. Each entry satisfies $\Sigma_{ij}=a(i-j)$, implying that the correlation between two ports depends only on their index separation and is independent of their absolute positions along the array. This shift-invariant (Toeplitz) structure is a defining characteristic of a wide-sense stationary process. 

This has motivated us to parameterize the spatial dependence through a finite-memory dynamic model. Specifically, we model $\{g_k\}_{k=1}^N$ as a stationary complex AR($p$) process \cite[Chapters~3.3--3.4]{BrockwellDavis}, \cite[Chapter~3]{BoxJenkins}
\begin{equation}\label{eq:ARp_def}
g_k = \underbrace{\sum_{i=1}^{p} \alpha_i g_{k-i}}_{p\text{-memory depth}} + \varepsilon_k,
\end{equation}
where the weight coefficients $\{\alpha_1,\dots,\alpha_p\}$ characterize how each port is linearly predicted from its $p$ preceding neighbors, thereby encoding the spatial correlation strength and decay behavior along the array. Hence, the AR coefficients quantify the effective correlation range and directional coupling among adjacent ports. Moreover, the term $\{\varepsilon_k\}$ is a temporally white circularly symmetric complex Gaussian innovation process
\begin{equation}\label{eq:innov}
\varepsilon_k \sim \mathcal{CN}(0,\sigma_\varepsilon^2),~\mathbb{E}\{\varepsilon_k \varepsilon_{k-\ell}^\ast\}=0,\ \ell\neq 0,
\end{equation}
independent of $\{g_{k-1},\dots,g_{k-p}\}$, {\em for compensating the factors that cannot be linearly predicted from its $p$-memory neighborhood}, such as uncorrelated spatial fluctuation injected at each port. Overall, the model parameters are $\{\alpha_1,\dots,\alpha_p,\sigma_\varepsilon^2\}$, which together fully specify the second-order statistics of the process through a finite-order state evolution mechanism.

We determine the AR($p$) parameters by matching the low-lag\footnote{We use the term {\em low-lag} as a short phrase for indicating the correlation between adjacent or very close ports in the array. As the distance between the ports increases (i.e., as the lag increases), the correlation generally weakens.}  spatial correlation of the Clarke's model \eqref{eq.Clarke}, and define
\begin{equation}\label{eq:target_r}
r(\ell) \triangleq \mathbb{E}\{ g_k g_{k-\ell}^\ast \}.
\end{equation}
Since $\boldsymbol{\Sigma}$ is Toeplitz, one could observe that
\begin{equation}
\mathbb{E}\{ g_k g_{k-\ell}^\ast \}= \sigma^2 a(\ell),
\end{equation}
which depends only on the lag $\ell$. To approximate the Clarke's model in the second-order sense, we therefore enforce correlation matching for the first $p$ lags (ports):
\begin{equation}\label{eq.c_autoco}
r(\ell) = \sigma^2 a(\ell),~\ell = 0,1,\dots,p.
\end{equation}

Then for the AR($p$) process defined in \eqref{eq:ARp_def}, we obtain a set of orthogonality conditions by correlating the recursion with past samples. Specifically, multiplying \eqref{eq:ARp_def} by $g_{k-\ell}^\ast$ and taking expectations gives, for $\ell \ge 1$,
\begin{align}
r(\ell)
	&= \sum_{i=1}^{p}\alpha_i\,\mathbb{E}\{g_{k-i}g_{k-\ell}^\ast\}+ \mathbb{E}\{\varepsilon_k g_{k-\ell}^\ast\}. \label{eq:YW_deriv}
\end{align}
Under the standard assumption that $\varepsilon_k$ is white and uncorrelated with $\{g_{k-\ell}\}_{\ell\ge 1}$ by \eqref{eq:innov}, the last term in \eqref{eq:YW_deriv} vanishes. Moreover, wide-sense stationarity implies $\mathbb{E}\{g_{k-i}g_{k-\ell}^\ast\}=r(\ell-i)$, yielding for $\ell=1,\dots,p$,
\begin{equation}\label{eq:YW_scalar}
r(\ell) = \sum_{i=1}^{p}\alpha_i r(\ell-i),
\end{equation}
which are precisely the complex Yule-Walker normal equations \cite{Yule-Walker1,Yule-Walker2}, enforcing that the one-step linear predictor $\sum_{i=1}^{p}\alpha_i g_{k-i}$ matches $g_k$ in the mean-square sense, i.e., the prediction error is orthogonal to the span of the past $p$ samples.

\begin{algorithm}[t!]
\caption{AR($p$) Gauss-Markov Correlation Modeling}\label{alg:ARp_GM}
{\footnotesize
\begin{algorithmic}[1]
		\STATE \textbf{Input:} AR order $p$, Clarke generator $a(\ell)$ in \eqref{eq.Clarke}, variance $\sigma^2$
		\STATE \textbf{Output:} AR coefficients $\boldsymbol{\alpha}$, innovation variance $\sigma_\varepsilon^2$
		
		\STATE Low-lag matching \eqref{eq.c_autoco}:
		$r(\ell)=\sigma^2 a(\ell)$, $\ell=0,\dots,p$;
		
		\STATE Form Yule--Walker system \eqref{eq:R_toeplitz}:
		construct $\boldsymbol{R},\boldsymbol{r}$;
		
		\STATE Solve AR parameters \eqref{eq:YW_matrix}:
		$\boldsymbol{R}\boldsymbol{\alpha}=\boldsymbol{r}$;
		
		\STATE Innovation variance \eqref{eq:sigeps_closed}:
		$\sigma_\varepsilon^2 = r(0)-\boldsymbol{\alpha}^{\mathsf H}\boldsymbol{r}$;
		
		\RETURN $\boldsymbol{\alpha},\sigma_\varepsilon^2$
\end{algorithmic}}
\end{algorithm}

To write \eqref{eq:YW_scalar} in a more compact form, define vectors $\boldsymbol{\alpha},~\boldsymbol{r}$ and Toeplitz matrix $\boldsymbol{R}$ as
\begin{equation}\label{eq:R_toeplitz}
\left\{\begin{aligned}
\boldsymbol{\alpha} &= [\alpha_1,\dots,\alpha_p]^{\mathsf T},\\
\boldsymbol{r} &= [r(1),\dots,r(p)]^{\mathsf T},\\
\boldsymbol{R}& =
\begin{pmatrix}
r(0) & r(1) & \cdots & r(p-1)\\
r(1)^\ast & r(0) & \cdots & r(p-2)\\
\vdots & \vdots & \ddots & \vdots\\
r(p-1)^\ast & r(p-2)^\ast & \cdots & r(0)
\end{pmatrix}.
\end{aligned}\right.
\end{equation}
Hence, the AR($p$)-system in \eqref{eq:YW_scalar} can be written compactly as
\begin{equation}\label{eq:YW_matrix}
\boldsymbol{R}\boldsymbol{\alpha}=\boldsymbol{r}.
\end{equation}
With the autocorrelation values in \eqref{eq.c_autoco}, the AR coefficients are obtained as $\boldsymbol{\alpha}=\boldsymbol{R}^{-1}\boldsymbol{r}$. Under this construction, the AR($p$) process reproduces exactly the first $p$ correlation lags of the Clarke's model, while higher-order correlations are determined by the finite-order recursion. As a consequence, this provides a finite-order, covariance-matching approximation of the original Toeplitz Gaussian process.

After finding solutions to $\boldsymbol{\alpha}$ via \eqref{eq:YW_matrix}, we now explain how to calculate the innovation variance $\sigma_\varepsilon^2$ in \eqref{eq:innov}. Starting from the AR($p$) definition in \eqref{eq:ARp_def} and rearranging the model as $\varepsilon_k = g_k-\sum_{i=1}^{p}\alpha_i g_{k-i}$. Since $\varepsilon_k \sim \mathcal{CN}(0,\sigma_\varepsilon^2)$, by definition of variance calculation, we can calculate $\sigma_\varepsilon^2$ by
\begin{equation}
\sigma_\varepsilon^2 \triangleq \mathbb{E}\{|\varepsilon_k|^2\}=\mathbb{E}\left\{\left|g_k-\sum_{i=1}^{p}\alpha_i g_{k-i}\right|^2\right\}.
\end{equation}
Expanding the squared magnitude yields
\begin{align}\label{eq: variance_expand}
	\sigma_\varepsilon^2
	&=\mathbb{E}\left\{\left(g_k-\sum_{i=1}^{p}\alpha_i g_{k-i}\right)
	\left(g_k^\ast-\sum_{j=1}^{p}\alpha_j^\ast g_{k-j}^\ast\right)\right\}\nonumber\\
	&=\mathbb{E}\{g_k g_k^\ast\}
	-\sum_{j=1}^{p}\alpha_j^\ast \mathbb{E}\{g_k g_{k-j}^\ast\}
	-\sum_{i=1}^{p}\alpha_i \mathbb{E}\{g_{k-i} g_k^\ast\}\nonumber\\
	&\quad\quad+\sum_{i=1}^{p}\sum_{j=1}^{p}\alpha_i\alpha_j^\ast \mathbb{E}\{g_{k-i} g_{k-j}^\ast\}.
\end{align}
Using the definition $r(\ell)=\mathbb{E}\{g_k g_{k-\ell}^\ast\}$ and the correlation stationarity, the four terms can be respectively calculated as
$\mathbb{E}\{g_k g_k^\ast\}=r(0)$, $\mathbb{E}\{g_k g_{k-j}^\ast\}=r(j)$, $\mathbb{E}\{g_{k-i} g_k^\ast\}=r(-i)=r(i)^\ast$, and $\mathbb{E}\{g_{k-i} g_{k-j}^\ast\}=r(j-i)$. Substituting these results into \eqref{eq: variance_expand}, we obtain
\begin{equation}\label{eq:sigeps_expand_again}
\sigma_\varepsilon^2= r(0)-\sum_{j=1}^{p}\alpha_j^\ast r(j)-\sum_{i=1}^{p}\alpha_i r(i)^\ast+\sum_{i=1}^{p}\sum_{j=1}^{p}\alpha_i\alpha_j^\ast r(j-i). 
\end{equation}

To simplify this, we first introduce the Toeplitz correlation matrix $\boldsymbol{R}$ and vector $\boldsymbol{r}$ from \eqref{eq:R_toeplitz} and \eqref{eq:YW_matrix}. Then we have
\begin{equation}\label{eq:identity1}
\left\{\begin{aligned}
\boldsymbol{\alpha}^{\mathsf H}\boldsymbol{r}&=\sum_{j=1}^{p}\alpha_j^\ast r(j),\\
\boldsymbol{\alpha}^{\mathsf H}\boldsymbol{R}\boldsymbol{\alpha}&=\sum_{i=1}^{p}\sum_{j=1}^{p}\alpha_i\alpha_j^\ast r(j-i).
\end{aligned}\right.
\end{equation}
Since $\boldsymbol{\alpha}$ satisfies the Yule-Walker model $\boldsymbol{R}\boldsymbol{\alpha}=\boldsymbol{r}$ in \eqref{eq:YW_matrix},
left-multiplying by $\boldsymbol{\alpha}^{\mathsf H}$ yields a useful identity:
\begin{equation}\label{eq:identity2}
\boldsymbol{\alpha}^{\mathsf H}\boldsymbol{R}\boldsymbol{\alpha}=\boldsymbol{\alpha}^{\mathsf H}\boldsymbol{r}.
\end{equation}
Substituting \eqref{eq:identity1} and \eqref{eq:identity2} into \eqref{eq:sigeps_expand_again}, we get
\begin{align}\label{eq.indenties_cancel}
	-\boldsymbol{\alpha}^{\mathsf H}\boldsymbol{r}-\big(\boldsymbol{\alpha}^{\mathsf H}\boldsymbol{r}\big)^\ast
	+\boldsymbol{\alpha}^{\mathsf H}\boldsymbol{R}\boldsymbol{\alpha}
	&=
	-\boldsymbol{\alpha}^{\mathsf H}\boldsymbol{r}-\big(\boldsymbol{\alpha}^{\mathsf H}\boldsymbol{r}\big)^\ast
	+\boldsymbol{\alpha}^{\mathsf H}\boldsymbol{r}\nonumber\\
	&=
	-\big(\boldsymbol{\alpha}^{\mathsf H}\boldsymbol{r}\big)^\ast.
\end{align}

This cancellation is a direct consequence of the Yule-Walker normal equations and reflects the orthogonality principle: the innovation $\varepsilon_k$ is uncorrelated with the past $p$ samples of the process. As a result, plugging \eqref{eq.indenties_cancel} into \eqref{eq:sigeps_expand_again}, the innovation variance admits the closed-form expression
\begin{equation}\label{eq:sigeps_closed}
\sigma_\varepsilon^2=r(0)-\boldsymbol{\alpha}^{\mathsf H}\boldsymbol{r}.
\end{equation}

\begin{figure}[]
\centering
\includegraphics[width=\columnwidth]{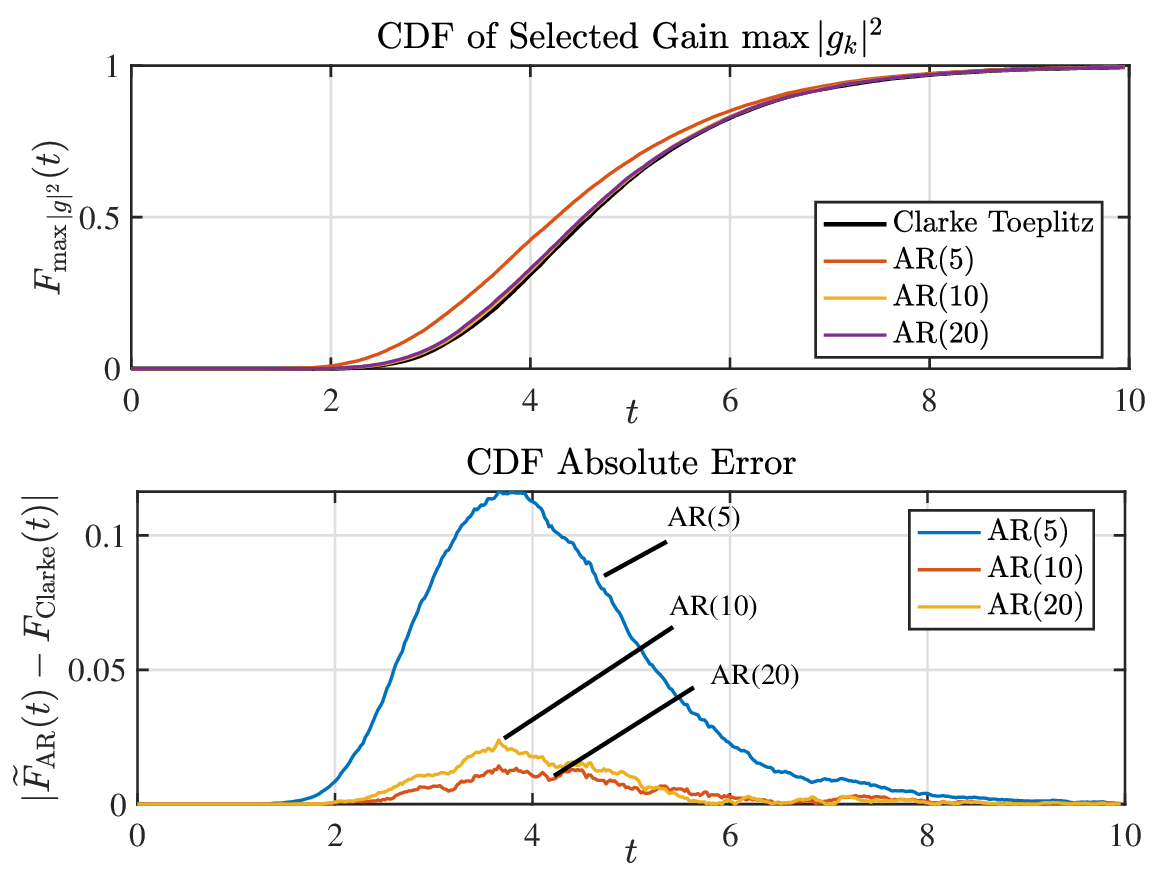}
\caption{Illustration of the proposed AR($p$) Gauss-Markov correlation modeling under different order coefficients $p\in \{5,10,20\}$ with $N=200$, $W=5$, burn-in length $B=5N$, indicating there is an optimal AR order under fixed $(W,N)$. All CDFs are generated via $3\times 10^4$ Monte-Carlo samples.}\label{fig:CDF_Different_Orders}
\vspace{-5mm}
\end{figure}

In summary, using \eqref{eq:R_toeplitz}, \eqref{eq:YW_matrix} and \eqref{eq:sigeps_closed}, the proposed correlation modeling through the AR($p$) Gauss-Markov process is completed. We summarize the approximation procedures in Algorithm~\ref{alg:ARp_GM}. Fig.~\ref{fig:CDF_Different_Orders} illustrates an example with different order coefficients $p\in \{5,10,20\}$ under $N=200$, $W=5$, burn-in length $B=5N$ (to be explained later). The effectiveness of the proposed correlation modeling is validated and under fixed $(W,N)$, the order of AR($p$) should be properly determined, which will be discussed in Section \ref{sec:p_selection}.

\subsection{Optimal AR Order $p^\star$ for Fixed $(W,N)$}\label{sec:p_selection}
Define the AR polynomial in the $z$-domain
\begin{equation}\label{eq:AR_poly}
A(z)=1-\sum_{i=1}^p \alpha_i z^{-i}.
\end{equation}
A standard stability condition is that all the roots of $A(z)$ lie inside the unit circle. When $p$ is too large, the resulting near-unstable dynamics distort long-range dependence and extreme-value statistics \cite{BrockwellDavis,BoxJenkins}. Given {\em fixed}\footnote{When \( N \) is fixed, increasing \( p \) (the AR order) leads to overfitting and poor estimation accuracy. This occurs because as \( p \to \infty \), the number of model parameters grows unbounded while \( N \) remains finite. In this case, the model cannot capture the underlying process accurately, leading to high variance in the parameter estimates. However, if \( N \to \infty \) as well, the system becomes capable of estimating an increasing number of parameters with arbitrarily high precision. This is because the data sample grows sufficiently large to support the increasing number of parameters, allowing the model to converge to the true underlying process and yield perfect accuracy in the limit.} $(W,N)$, the correlation sequence $a(\ell)$ is not exactly representable by any finite-order AR model. Increasing $p$ improves low-lag correlation matching, but an excessively large $p$ may induce near-instability and distort tail behavior. Hence, there exists an optimal finite order $p^\star$ that best balances correlation fidelity and stability under fixed $(W,N)$. In the following, we explain how to properly set the order of AR by a step-wise example of approximating the distribution of random variable $g_{\max}=\max_{1\le k\le N}|g_k|^2$.

{\em Target to be Matched}---The quantity of interest here is the FAS selection gain, defined as
\begin{equation}
g_{\max}=\max_{1\le k\le N}|g_k|^2.
\end{equation}
Let $F(t)$ denote the CDF of $g_{\max}$ under the true Toeplitz-Clarke model and let $\widetilde{F}(t)$ denote the CDF under the AR($p$) approximation. The optimal order $p^*$ with a tolerable range $[1,p_{\max}]$ is defined as the one whose induced distribution of $g_{\max}$ is closest to the true one in the sense of
\begin{align}
D(p)&=\sup_{t\ge0}\big|F(t)-\widetilde{F}(t)\big|, \label{eq:KS_def}\\
p^\star&=\arg\min_{p\in\{1,\dots,p_{\max}\}}D(p),\label{eq:KS_def_order}
\end{align}
where \eqref{eq:KS_def} captures the approximation under worst case and \eqref{eq:KS_def_order} outputs the optimal order within $p\in\{1,\dots,p_{\max}\}$.

{\em Practical Evaluation Procedure}---In practice, both $F$ and $\widetilde{F}$ are evaluated via Monte Carlo experiments. The procedure consists of the following four steps:
\begin{enumerate}
\item \textbf{Reference distribution}---Generate $L$ samples $\boldsymbol g^{(i)}\sim \mathcal{CN}(\boldsymbol 0,\boldsymbol\Sigma)$ with $\Sigma_{ij}=\sigma^2 a(i-j)$. Then compute samples $g_{\max}^{(i)}=\max_{1\le k\le N}|g_k^{(i)}|^2$, for $i=1,\dots,L,$ and construct the empirical CDF $F$.
\item \textbf{AR($p$) fitting}---For each candidate $p$, set target lags $r(\ell)=\sigma^2 a(\ell)$ for $\ell=0,\dots,p$. Solve the Yule-Walker system to obtain $\{\alpha_i\}$ and $\sigma_\varepsilon^2$ by Algorithm~\ref{alg:ARp_GM}.
\item \textbf{AR($p$) simulation}---Using the fitted parameters, generate $L$ independent AR($p$) realizations of length $N+B$, discard the first $B$ samples (burn-in), and then compute $g_{\max}^{(i)}=\max_{1\le k\le N}|g_k^{(i)}|^2$. From the samples, construct the empirical CDF $\widetilde{ F}$.
\item \textbf{Evaluation and order selection}---Compute \eqref{eq:KS_def} and choose $p^\star$ by \eqref{eq:KS_def_order}.
\end{enumerate}

Since $p$ affects the computational complexity, selecting the appropriate approximation order within a tolerable range is the goal in this section. Also, the largest memory depth should be no more than the total number of ports, i.e., $p^{\star}\le N$.

\vspace{-2mm}
\subsection{Random Realizations via AR($p$) Gauss-Markov Modeling}\label{subsec:ARp_generation}
Here, we explain how to generate randomized samples by using the correlation approximation through the AR($p$) Gauss-Markov correlation modeling introduced in Section \ref{subsec:ARp_model}. Once the AR($p$) parameters $\{\alpha_i,\sigma_\varepsilon^2\}$ are determined, we generate an $N$-port channel realization by simulating a \emph{stationary} AR($p$) trajectory and then extracting a length-$N$ block. The key practical issue is that AR recursions require \emph{initial conditions} to offer randomized samples with stable distribution. If one starts from arbitrary initial states (e.g., zeros), the first part of the simulated sequence is generally \emph{not} distributed according to the stationarity law implied by $\{\alpha_i,\sigma_\varepsilon^2\}$. This initial transient may bias statistics that are sensitive to tails/extremes (e.g., $\max_k |g_k|^2$). Hence, we employ a \emph{burn-in} period \cite{BrockwellDavis,BoxJenkins}.

Burn-in \cite{BrockwellDavis2016} means that we simulate the AR($p$) recursion for $B+N$ samples, discard these first $B$ samples, and keep the subsequent $N$ samples as the effective stationary block used for the FAS ports. Let $B$ denote the burn-in length. We generate a sequence of length $B+N$ from the AR($p$) recursion \eqref{eq:ARp_def} and then set $\boldsymbol{g}_u^{(\mathrm{AR})}=[g_1,\dots,g_N]^{\mathsf T}$ from the final $N$ samples after burn-in. The full autoregression procedures are:

\begin{itemize}
\item \textbf{Step 1:} Choose the port length $N$ and a burn-in length $B$. One should simulate a total of $B+N$ samples.
\item \textbf{Step 2:} Provide an initial state $(g_{0},g_{-1},\dots,g_{-p+1})$. A simple choice is $g_{0}=g_{-1}=\cdots=g_{-p+1}=0$, I which is \emph{not} stationary but is acceptable because burn-in will remove the transient.
\item \textbf{Step 3:} Draw independent and identically distributed (i.i.d.) innovations $\varepsilon_1,\varepsilon_2,\dots,\varepsilon_{B+N}\stackrel{\mathrm{i.i.d.}}{\sim}\mathcal{CN}(0,\sigma_\varepsilon^2)$. These $\varepsilon_k$ are the \emph{only} new randomness among all steps and all $g_k$ randomness propagates from the recursion.
\item \textbf{Step 4:} For $k=1,2,\dots,B+N$, compute (\ref{eq:ARp_def}). This produces a trajectory $\{g_{-p+1},\dots,g_0,g_1,\dots,g_{B+N}\}$.
\item \textbf{Step 5:} Discard the first $B$ generated samples $\{g_1,\dots,g_B\}$. Intuitively, this removes dependence on the arbitrary initialization and yields samples that are close to the stationary distribution implied by $\{\alpha_i,\sigma_\varepsilon^2\}$.
\item \textbf{Step 6:} Keep the next $N$ samples and define $\boldsymbol{g}_u^{(\mathrm{AR})}\triangleq[g_{B+1},g_{B+2},\dots,g_{B+N}]^{\mathsf T}$. For notational simplicity, we relabel $(g_{B+1},\dots,g_{B+N})$ as $(g_1,\dots,g_N)$ when using the vector as an $N$-port channel realization.
\end{itemize}

This construction yields a finite-memory approximation to the original Toeplitz Gaussian vector $\boldsymbol{g}_u\sim\mathcal{CN}(\boldsymbol{0},\boldsymbol{\Sigma})$ while preserving the dominant low-lag correlation structure through the AR($p$)-fitted parameters $\{\alpha_1,\dots,\alpha_p,\sigma_\varepsilon^2\}$ with necessary burn-in step in practice for stable sample generations.

\begin{figure}[]
\centering
\includegraphics[width=0.85\columnwidth]{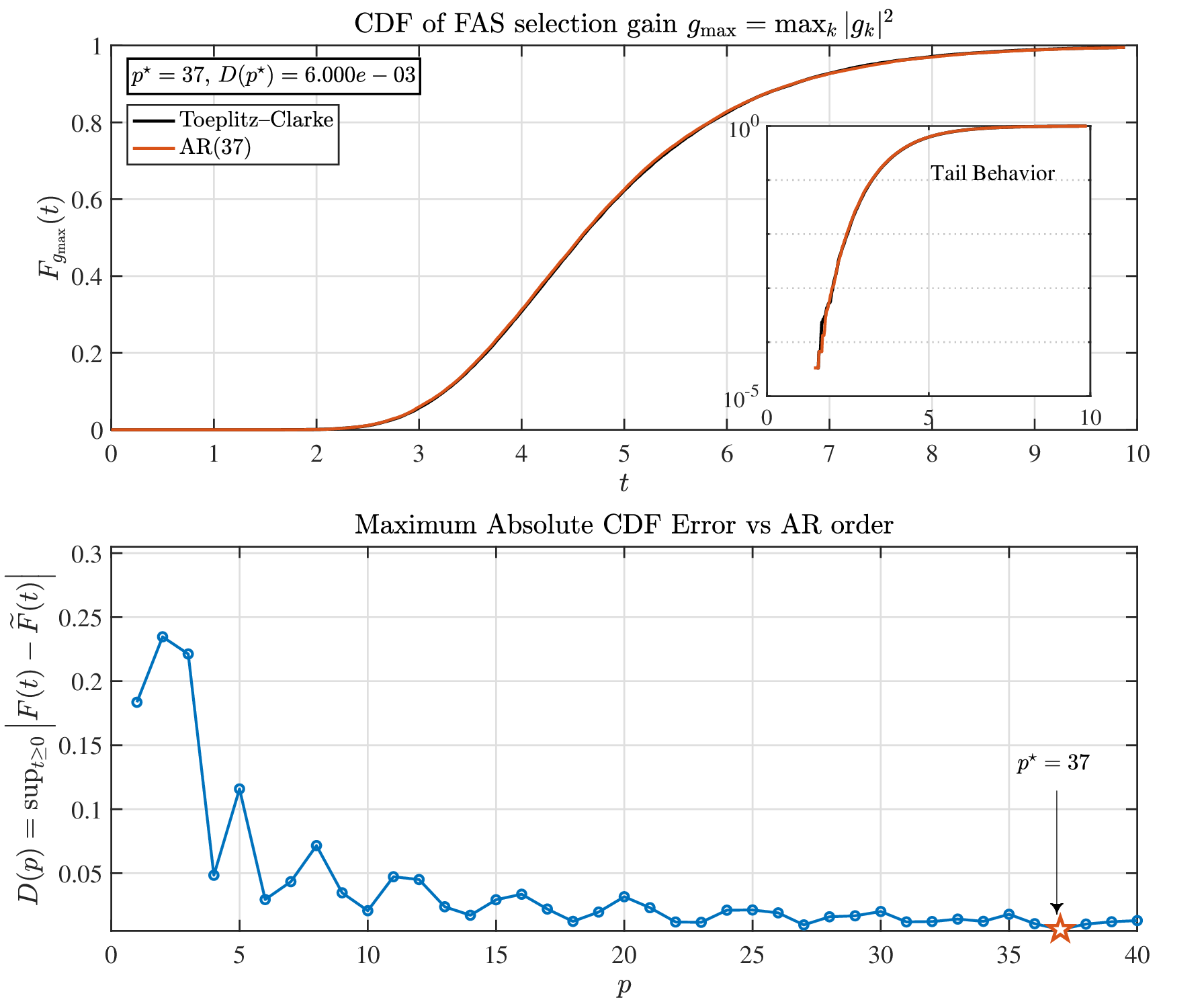}
\caption{Illustration of the maximum error of the proposed AR($p$) Gauss-Markov correlation modeling under different order coefficients $p\in[1,40]$ with $N=200$, $W=5$ and burn-in length $B=5N$, where the maximum CDF approximation error is $6\times 10^{-3}$ with $p^{*}=37$ within the range of $p\in[1,40]$. All CDFs are generated via $3\times 10^4$ Monte-Carlo samples.}\label{fig:CDF_Different_Error_Orders}
\vspace{-3mm}
\end{figure}

Fig.~\ref{fig:CDF_Different_Error_Orders} illustrates the maximum absolute CDF approximation error under order range $p\in [1,40]$ where $p^{*}$ is determined by Section \ref{sec:p_selection} with $N=200$, $W=5$ and burn-in length $B=5N$. Clearly, the approximation remains accurate in terms of CDF {\em tendency} and the {\em tail behavior}, validating the proposed correlation approximation mindset via generative modeling viewpoint instead of descriptive covariance approximation.

\vspace{-3mm}
\section{Analytical Distribution of the FAS Selection Gain under Toeplitz-Clarke and AR($p$)}\label{subsec:ARmax_dist}
Thus far, we have demonstrated the effectiveness of the AR($p$) approximation for correlation modeling and sample generation, this section establishes its fundamental analytical advantages. Specifically, we derive a structured expression for the CDF of the FAS selection gain by leveraging the Markov properties embedded within the AR($p$) recursion. Under the exact Toeplitz-Clarke model, the spatial port vector is given by $\boldsymbol{g} = [g_1, \dots, g_N]^{\mathsf{T}} \sim \mathcal{CN}(\boldsymbol{0}, \boldsymbol{\Sigma})$, where $\Sigma_{ij} = \sigma^2 a(i-j)$ defines the dense spatial correlation. The quantity of paramount interest for evaluating system performance is the selection gain, i.e., $g_{\max} = \max_{1 \le k \le N} |g_k|^2$. The CDF of \(g_{\max}\), which underpins the outage probability analysis, is formulated as the joint probability that all port magnitudes are less than some threshold $t$, i.e.,
\begin{equation}
F_{\mathrm{T}}(t) = \Pr(g_{\max} \le t) = \Pr(|g_1|^2 \le t, \dots, |g_N|^2 \le t).
\end{equation}
Mathematically, evaluating this CDF requires calculating the integral of the joint PDF over an $N$-dimensional truncated complex region $ \mathcal{D}(t)$, written as
\begin{equation}\label{eq:FT_integral}
F_{\mathrm{T}}(t) = \int_{\mathcal{D}(t)} f_{\boldsymbol{g}}(\boldsymbol{z})\, d\boldsymbol{z},
\end{equation}
where $ f_{\boldsymbol{g}}(\boldsymbol{z}) $ is the multivariate Gaussian density
\begin{equation}
f_{\boldsymbol{g}}(\boldsymbol{z}) = \frac{1}{\pi^N \det(\boldsymbol{\Sigma})} \exp\left(- \boldsymbol{z}^{\mathsf{H}} \boldsymbol{\Sigma}^{-1} \boldsymbol{z}\right),
\end{equation}
and the integration domain is defined by the Cartesian product $\mathcal{D}(t) = \{\boldsymbol{z} \in \mathbb{C}^N : |z_k|^2 \le t \ ,\forall k\}$. 

The fundamental difficulty in evaluating \eqref{eq:FT_integral} arises from the quadratic form $\boldsymbol{z}^{\mathsf{H}} \boldsymbol{\Sigma}^{-1} \boldsymbol{z}$, which introduces dense coupling across all $N$ coordinates. Consequently, this high-dimensional integral lacks a tractable closed-form solution. To circumvent this analytical bottleneck, we demonstrate how the proposed AR($p$) model transforms the globally coupled integration into a sequential, low-dimensional recursive process. The conceptual comparison between the exact model and the proposed approximation is summarized in Table~\ref{tab:tc_vs_arp}.

\begin{table*}[]
\caption{Toeplitz-Clarke vs.~AR($p$) approximation for the CDF of $g_{\max}=\max_{1\le k\le N}|g_k|^2$.}\label{tab:tc_vs_arp}
\centering
\footnotesize
\renewcommand{\arraystretch}{1.15}
\resizebox{.85\linewidth}{!}{
\begin{tabularx}{\linewidth}{lXX}
		\toprule
		\textbf{Aspect} & \textbf{Toeplitz-Clarke (before approximation)} & \textbf{AR($p$) Gauss-Markov (after approximation)} \\
		\midrule
		Underlying model
		&
		$\boldsymbol g=[g_1,\dots,g_N]^{\mathsf T}\!\sim\!\mathcal{CN}(\boldsymbol 0,\boldsymbol\Sigma)$,  
		$\Sigma_{ij}=\sigma^2 a(i-j)$
		&
		$g_k=\sum_{i=1}^p\alpha_i g_{k-i}+\varepsilon_k$,  
		$\varepsilon_k\sim\mathcal{CN}(0,\sigma_\varepsilon^2)$
		\\[2mm]
		Target CDF
		&
		$F_{\mathrm T}(t)=\Pr(g_{\max}\le t)$
		&
		$F_p(t)=\Pr(g_{\max}\le t)$
		\\[2mm]
		CDF representation
		&
		$F_{\mathrm T}(t)=\int_{\mathcal D(t)} f_{\boldsymbol g}(\boldsymbol z)\,d\boldsymbol z$,  
		$\mathcal D(t)=\{\boldsymbol z:|z_k|^2\le t,\forall k\}$
		&
		$F_p(t)=\int_{\mathbb C^p}\phi_N(\boldsymbol s)\,d\boldsymbol s$,  
		$\phi_{k+1}=\mathbf 1\{|z_1'|^2\le t\}\!\int \phi_k f(\boldsymbol s'|\boldsymbol s)d\boldsymbol s$
		\\[2mm]
		Where difficulty comes from
		&
		Dense coupling in $\boldsymbol z^{\mathsf H}\boldsymbol\Sigma^{-1}\boldsymbol z$  
		+ per-coordinate truncation in $\mathcal D(t)$
		&
		Truncation handled sequentially via Markov state;  
		effective dimension depends on $p$ rather than $N$
		\\[2mm]
		\bottomrule
\end{tabularx}}
\vspace{-3mm}
\end{table*}

\vspace{-2mm}
\subsection{AR($p$) Model and Markov Factorization}\label{subsec:ARp_markov_factorization}
By approximating the channel $\{g_k\}_{k=1}^N$ with a stationary AR($p$) process $g_k=\sum_{i=1}^{p}\alpha_i g_{k-i}+\varepsilon_k$, the structural dependence is localized. The CDF of the selection gain under the AR($p$) framework can be expressed as
\begin{equation}\label{eq:Fp_def}
F_p(t)\triangleq \Pr(g_{\max}\le t)=\Pr\left(\bigcap_{k=1}^{N}\{|g_k|^2\le t\}\right).
\end{equation}
For a given threshold $t\ge 0$, we introduce the \emph{non-exceedance} event up to port index $k$
\begin{equation}\label{eq:Ek_def}
\mathcal{E}_k(t)\triangleq \{|g_1|^2\le t,\dots,|g_k|^2\le t\}.
\end{equation}
In the context of stochastic processes, $\mathcal{E}_k(t)$ represents the survival event (i.e., the process remains strictly within the threshold boundaries up to time $k$). Thus, the CDF simplifies to the terminal survival probability $F_p(t)=\Pr(\mathcal{E}_N(t))$.

To exploit the sequential structure, we define the $p$-dimensional lifted state vector
\begin{equation}\label{eq:state_def}
\boldsymbol{s}_k \triangleq [g_k, g_{k-1}, \dots, g_{k-p+1}]^{\mathsf T}\in\mathbb{C}^p,
\end{equation}
which forms a time-homogeneous first-order Markov process. Conditioned on the state $\boldsymbol{s}_k = [z_1,\dots,z_p]^{\mathsf T}$, the subsequent sample $g_{k+1}$ is a complex Gaussian random variable
\begin{equation}\label{eq:cond_gauss}
g_{k+1}\mid \boldsymbol{s}_k\sim \mathcal{CN}(\mu(\boldsymbol{s}_k),\sigma_\varepsilon^2),~\mbox{where } \mu(\boldsymbol{s}_k)=\sum_{i=1}^p \alpha_i z_i,
\end{equation}
while the remaining historical components shift deterministically (i.e., $z_2'=z_1,\dots,z_p'=z_{p-1}$). This state-space representation is a foundational tool in AR modeling \cite{DurbinKoopman2012}.

Rather than tracking the survival probability $\Pr(\mathcal{E}_k(t))$ directly, it is analytically advantageous to propagate the joint PDF of the Markov state $\boldsymbol{s}_k$ and the non-exceedance event $\mathcal{E}_k(t)$ simultaneously. We define this joint density as
\begin{equation}
	\phi_k(\boldsymbol{s})
	\triangleq
	f_{\boldsymbol{s}_k,\mathcal{E}_k(t)}(\boldsymbol{s}).
	\label{eq:phi_def}
\end{equation}
Equivalently, this un-normalized density can be explicitly expanded as the product of the unconstrained state density and an indicator function enforcing the threshold conditions
\begin{equation}
	\phi_k(\boldsymbol{s})
	=
	f_{\boldsymbol{s}_k}(\boldsymbol{s})
	\;
	\mathbf{1}\{|z_1|^2\le t,\dots,|z_{\min(k,p)}|^2\le t\}.
	\label{eq:phi_expanded}
\end{equation}
Crucially, the total integrated mass of $\phi_k(\boldsymbol{s})$ over the state space $\mathbb{C}^p$ exactly equals the survival probability $\Pr(\mathcal{E}_k(t))$.

Let $f(\boldsymbol{s}'|\boldsymbol{s})$ denote the one-step transition density from state $\boldsymbol{s}_k=\boldsymbol{s}$ to $\boldsymbol{s}_{k+1}=\boldsymbol{s}'$. Based on the AR($p$) recursion, this density factorizes into a stochastic generating term and a deterministic shifting term
\begin{equation}\label{eq:trans_density}
f(\boldsymbol{s}'|\boldsymbol{s})=f(z_1'|\boldsymbol{s})\prod_{j=2}^{p}\delta(z_j'-z_{j-1}),
\end{equation}
where $\delta(\cdot)$ represents the Dirac delta function, and the conditional density of the new sample is
\begin{equation}\label{eq:trans_gauss}
f(z_1'|\boldsymbol{s})=\frac{1}{\pi\sigma_\varepsilon^2}\exp\!\left(-\frac{|z_1'-\mu(\boldsymbol{s})|^2}{\sigma_\varepsilon^2}\right).
\end{equation}

By the Chapman-Kolmogorov principle \cite{Chapman-Kolmogorov} and the law of total probability, $\phi_k$ can be recursively updated. Because the historical non-exceedance conditions are implicitly encoded within $\phi_k(\boldsymbol{s})$, the update rule from step $k$ to $k+1$ only needs to enforce the threshold condition on the newly generated sample $z_1'$. This yields the forward integration recursion
\begin{equation}\label{eq:phi_recursion}
\phi_{k+1}(\boldsymbol{s}')=\mathbf{1}\{|z_1'|^2\le t\}\int_{\mathbb{C}^p}\phi_k(\boldsymbol{s})f(\boldsymbol{s}'|\boldsymbol{s})d\boldsymbol{s},~k\ge p.
\end{equation}
Structurally, \eqref{eq:phi_recursion} governs a Feynman-Kac flow \cite{Feynman-Kac}, wherein the Markov transition kernel $f(\boldsymbol{s}'|\boldsymbol{s})$ interacts with a multiplicative potential (the indicator function) acting as an absorptive barrier. The desired target CDF is recovered by evaluating the remaining probability mass at the terminal stage $N$:
\begin{equation}\label{eq:Fp_from_phi}
F_p(t)=\Pr(\mathcal{E}_N(t))=\int_{\mathbb{C}^p}\phi_N(\boldsymbol{s})\,d\boldsymbol{s}.
\end{equation}
Overall, equations \eqref{eq:phi_recursion} and \eqref{eq:Fp_from_phi} provide an exact characterization of $F_p(t)$ under the AR($p$) assumption, successfully substituting the intractable $N$-dimensional integration with a sequence of tractable $p$-dimensional updates.

\vspace{-2mm}
\subsection{Particle-based Evaluation of the Markov Recursion}\label{subsec:smc_forward}
The recursive structure in \eqref{eq:phi_recursion} and \eqref{eq:Fp_from_phi} elegantly establishes analytical tractability, but evaluating the $p$-dimensional integrals analytically or via deterministic quadrature becomes computationally demanding when $F_p(t)$ is required across dense threshold grids. The theoretical objective can be recast as an expectation over the sequential indicator products
\begin{equation}\label{eq:F_indicator}
F_p(t)=\mathbb{E}\left[\prod_{k=1}^{N}\mathbf{1}\{|g_k|^2\le t\}\right].
\end{equation}

Direct Monte Carlo estimation of \eqref{eq:F_indicator} is computationally expensive. For small thresholds $t$ or large array sizes $N$, the joint non-exceedance becomes a rare event, and most sampled trajectories violate the constraints early, contributing zero to the expectation and leading to unacceptable sample waste.

To tackle this, we adopt a sequential Monte Carlo (SMC) particle filtering approach, following the Feynman-Kac flow of \eqref{eq:phi_recursion}. Instead of evaluating complete trajectories, SMC dynamically propagates a population of particles, continuously pruning those that violate the threshold and re-weighting the survivors to estimate the conditional survival probabilities.

For a fixed threshold $t$, we initialize $J$ particles $\{\boldsymbol{s}_k^{(j)}\}_{j=1}^J$ equipped with importance weights $\{w_k^{(j)}\}$. At transition step $k+1$, each particle state evolves by generating a new sample
\begin{equation}\label{eq:smc_prop}
g_{k+1}^{(j)}=\mu(\boldsymbol{s}_k^{(j)})+\varepsilon_{k+1}^{(j)},~\mbox{where }\varepsilon_{k+1}^{(j)}\sim\mathcal{CN}(0,\sigma_\varepsilon^2),
\end{equation}
accompanied by deterministic shifting of the historical components. The corresponding binary indicator for each particle is evaluated as $I_{k+1}^{(j)}(t) = \mathbf{1}\{|g_{k+1}^{(j)}|^2\le t\}$. The particle weights are then updated multiplicatively and normalized
\begin{equation}\label{eq:smc_weight}
\tilde w_{k+1}^{(j)}=w_k^{(j)} I_{k+1}^{(j)}(t),~\mbox{where }w_{k+1}^{(j)}=\frac{\tilde w_{k+1}^{(j)}}{\sum_{m=1}^{J}\tilde w_{k+1}^{(m)}}.
\end{equation}
Then the conditional survival probability for step $k+1$, given survival up to step $k$, can be reliably estimated by the weighted sum of the surviving particles
\begin{equation}\label{eq:ck_hat_new}
\hat c_{k+1}(t)=\sum_{j=1}^{J}w_k^{(j)} I_{k+1}^{(j)}(t).
\end{equation}
Analytically, the unconditional CDF, $F_p(t)$, is the product of these sequential conditional probabilities
\begin{equation}\label{eq:F_product}
F_p(t)=\prod_{k=1}^{N} \hat c_k(t).
\end{equation}

In practice, the numerical estimate is constructed by accumulating the logarithms of the conditional estimates, which inherently prevents arithmetic underflow, i.e.,
\begin{equation}
\left\{\begin{aligned}
\widehat{F}_p(t) &= \prod_{k=1}^{N} \hat c_k(t),\\
\log \widehat{F}_p(t) &= \sum_{k=1}^{N} \log \hat c_k(t).
\end{aligned}\right.
\end{equation}
To address particle degeneracy, a phenomenon where consecutive truncations concentrate all probability mass on a negligible fraction of particles, we track the effective sample size (ESS)
\begin{equation}\label{eq:ESS}
\mathrm{ESS}_k=\frac{1}{\sum_{j=1}^{J}(w_k^{(j)})^2}.
\end{equation}
If the ratio $\mathrm{ESS}_k/J$ drops below a specified threshold (e.g., $0.5$), a systematic resampling step is executed, duplicating highly-weighted particles, eliminating obsolete ones, and uniformly resetting the weights to $1/J$. The computational complexity of this particle-based evaluation is $\mathcal{O}(J N)$ per target threshold $t$ (augmented by a factor linear in $p$ for computing $\mu(\boldsymbol{s})$). This linear scaling strictly bypasses the dimensionality bottlenecks of the exact Toeplitz-Clarke integration.

\begin{figure}[]
\centering
\includegraphics[width=\columnwidth]{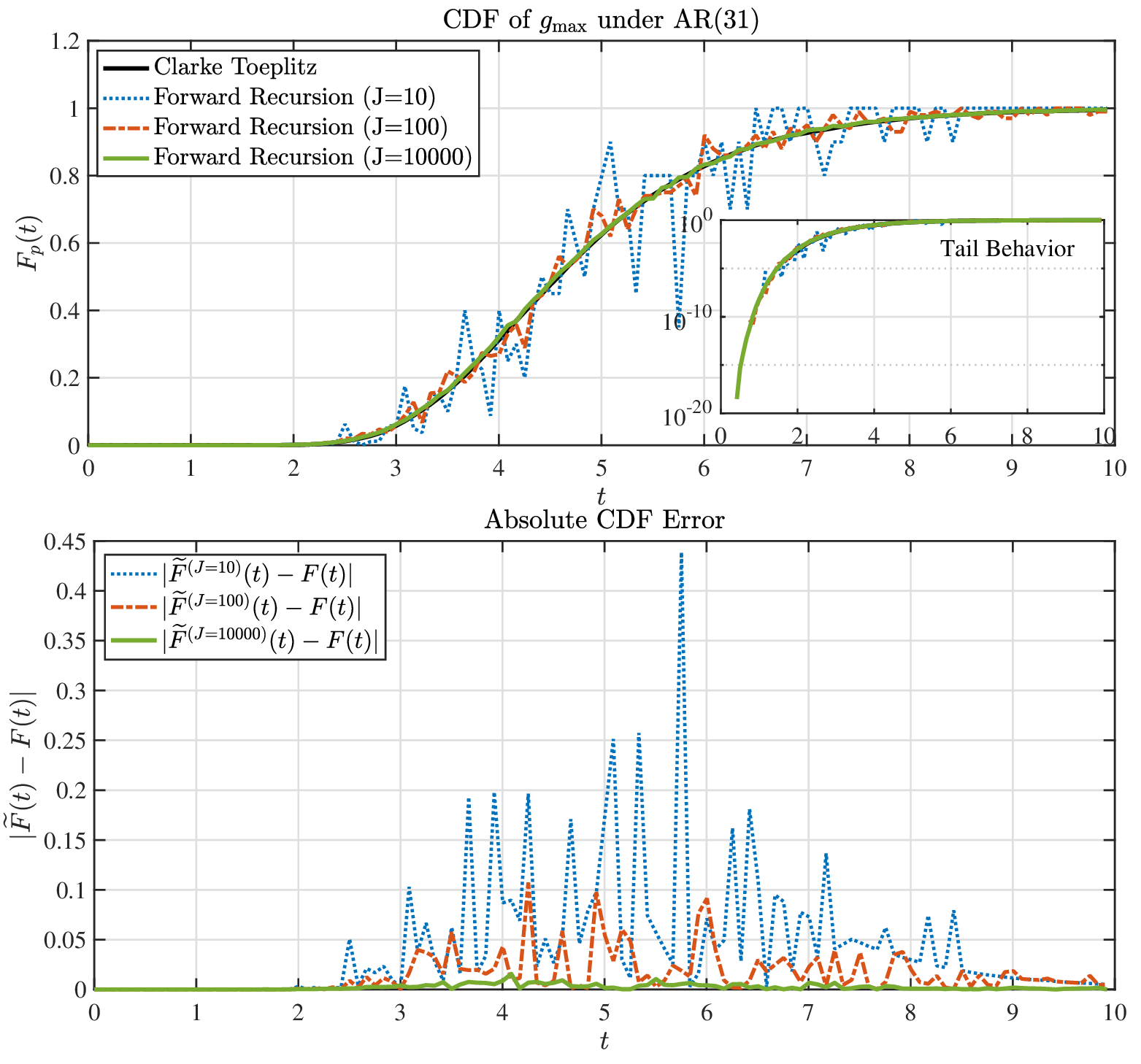}
\caption{Illustration of the analytically structured CDF by \eqref{eq:F_product} under the AR($p$) recursion modeling via particle-based evaluation ($N=200$, $W=5$, burn-in length $B=5N$).}\label{fig:Analytical_Partical_CDFs}
\vspace{-4mm}
\end{figure}
 
As validated by Fig.~\ref{fig:Analytical_Partical_CDFs}, increasing the particle swarm size $J$ yields highly consistent analytical approximations. As such, this confirms that evaluating the AR($p$) structured CDF provides a rigorous, scalable numerical methodology for extreme-value statistical analysis in FAS channels.

\vspace{-2mm}
\section{Channel Interpolation from Arbitrary \\Sparse Port Observations}\label{sec:channel_completion}
This section extends the AR($p$)-based correlation approximation beyond port-selection statistics to a practically useful inference task, i.e., \emph{channel reconstruction via interpolation}. Specifically, given $N$ ports, we assume that only an arbitrary subset of complex channel samples is observed or estimated. Our objective is to infer/reconstruct the remaining unobserved ports utilizing the correlation embedded within the model. Let $\mathcal O \subset \{1,\dots,N\}, ~ |\mathcal O|=M$ be the observed index set, with observed values $\{g_k\}_{k\in\mathcal O}$. The unobserved index set is
\begin{equation}
\mathcal U = \{1,\dots,N\}\setminus\mathcal O.
\end{equation}
Our goal is to reconstruct the unknown channel coefficients $\{g_k\}_{k\in\mathcal U}$ from the observations. In the sequel, we adopt the fitted AR($p$) model obtained from the correlation-matching stage in \eqref{eq:ARp_def}, which is linear and driven by Gaussian noise.

\vspace{-2mm}
\subsection{Interpolation via Joint Gaussian Distribution}\label{subsec:gaussian_feasibility}
Under the AR($p$) model, the stacked vector $\boldsymbol g$ is jointly circularly-symmetric complex Gaussian  where $\boldsymbol\Sigma\in\mathbb{C}^{N\times N}$ is the covariance matrix (either induced/approximated or the true one such as that in the Clarke's model). Reorder $\boldsymbol g$ as
\begin{equation}
	\boldsymbol g =
	\begin{bmatrix}
		\boldsymbol g_{\mathcal O} \\
		\boldsymbol g_{\mathcal U}
	\end{bmatrix},
\end{equation}
where $\boldsymbol g_{\mathcal O} = \{g_k\}_{k\in\mathcal O}$, $\boldsymbol g_{\mathcal U} = \{g_k\}_{k\in\mathcal U}$. The covariance matrix is partitioned accordingly as
\begin{equation}
	\boldsymbol\Sigma =
	\begin{bmatrix}
		\boldsymbol\Sigma_{\mathcal O\mathcal O} &
		\boldsymbol\Sigma_{\mathcal O\mathcal U} \\
		\boldsymbol\Sigma_{\mathcal U\mathcal O} &
		\boldsymbol\Sigma_{\mathcal U\mathcal U}
	\end{bmatrix}.
\end{equation}
Since the joint distribution is Gaussian, the conditional distribution of $\boldsymbol g_{\mathcal U}$ given $\boldsymbol g_{\mathcal O}$ is also complex Gaussian:
\begin{equation}
	\boldsymbol g_{\mathcal U} \mid \boldsymbol g_{\mathcal O}
	\sim
	\mathcal{CN}
	\big(
	\boldsymbol\mu_{\mathcal U|\mathcal O},
	\boldsymbol\Sigma_{\mathcal U|\mathcal O}
	\big),
\end{equation}
where the conditional mean is given by
\begin{equation}
	\boldsymbol\mu_{\mathcal U|\mathcal O}
	=
	\boldsymbol\Sigma_{\mathcal U\mathcal O}
	\boldsymbol\Sigma_{\mathcal O\mathcal O}^{-1}
	\boldsymbol g_{\mathcal O},
	\label{eq:cond_mean_clean}
\end{equation}
and the conditional covariance is given by
\begin{equation}
	\boldsymbol\Sigma_{\mathcal U|\mathcal O}
	=
	\boldsymbol\Sigma_{\mathcal U\mathcal U}
	-
	\boldsymbol\Sigma_{\mathcal U\mathcal O}
	\boldsymbol\Sigma_{\mathcal O\mathcal O}^{-1}
	\boldsymbol\Sigma_{\mathcal O\mathcal U}.
	\label{eq:cond_cov_clean}
\end{equation}
Therefore, channel completion from arbitrary sparse observations is theoretically feasible and admits a {\em globally optimal solution} under the Gaussian prior via an MMSE estimator. The MMSE estimator equals the conditional mean:
\begin{equation}
		\widehat{\boldsymbol g}_{\mathcal U}^{\mathrm{MMSE}}
		=
		\boldsymbol\Sigma_{\mathcal U\mathcal O}
		\boldsymbol\Sigma_{\mathcal O\mathcal O}^{-1}
		\boldsymbol g_{\mathcal O},
	\label{eq:mmse_unobs_clean}
\end{equation}
whose corresponding error covariance is $\boldsymbol\Sigma_{\mathcal U|\mathcal O}$ in \eqref{eq:cond_cov_clean}.

Let $\mathbf S_{\mathcal O}\in\{0,1\}^{M\times N}$ be the selection matrix extracting observed entries, where there is only a single element valued to $1$ at each row. Thus, the port observed can be found as
\begin{equation}\label{eq.selection_matrix}
\boldsymbol g_{\mathcal O}=\mathbf S_{\mathcal O}\boldsymbol g.
\end{equation}
Assume potential noisy observations
\begin{equation}\label{eq:noisy_observe}
\boldsymbol y=\mathbf S_{\mathcal O}\boldsymbol g+\boldsymbol v,~\boldsymbol v\sim\mathcal{CN}(\boldsymbol 0,\sigma_v^2\boldsymbol I).
\end{equation}
Then the MMSE admits the closed form
\begin{equation}\label{eq:mmse_selection_clean}
\widehat{\boldsymbol g}^{\mathrm{MMSE}}=\boldsymbol\Sigma\mathbf S_{\mathcal O}^{\mathsf H}\left(\mathbf S_{\mathcal O}\boldsymbol\Sigma\mathbf S_{\mathcal O}^{\mathsf H}+\sigma_v^2\boldsymbol I\right)^{-1}\boldsymbol y.
\end{equation}
Equivalently, using index notation, we have
\begin{equation}
\widehat{\boldsymbol g}^{\mathrm{MMSE}}=\boldsymbol\Sigma(:,\mathcal O)\left(\boldsymbol\Sigma(\mathcal O,\mathcal O)+\sigma_v^2\boldsymbol I\right)^{-1}\boldsymbol y.
\end{equation}
Moreover, the conditional error covariance can be obtained as
\begin{equation}
\mathrm{Cov}(\boldsymbol g| \boldsymbol y)=\boldsymbol\Sigma-\boldsymbol\Sigma\mathbf S_{\mathcal O}^{\mathsf H}\left(\mathbf S_{\mathcal O}\boldsymbol\Sigma\mathbf S_{\mathcal O}^{\mathsf H}+\sigma_v^2\boldsymbol I\right)^{-1}\mathbf S_{\mathcal O}\boldsymbol\Sigma.
\end{equation}
Hence, for the unobserved portion, the theoretical normalized mean-square-error (NMSE) is given by
\begin{equation}\label{eq:nmse_clean}
\mathrm{NMSE}_{\mathcal U}=\frac{\mathrm{tr}\big(\mathrm{Cov}(\boldsymbol g_{\mathcal U}| \boldsymbol y)\big)}{\mathrm{tr}\big(\boldsymbol\Sigma_{\mathcal U\mathcal U}\big)}.
\end{equation}
Although the Gaussian conditioning formulae \eqref{eq:cond_mean_clean} and \eqref{eq:cond_cov_clean} are conceptually straightforward, their implementation entails substantial computational cost. Specifically, it requires:
\begin{itemize}
\item Explicit construction of $\boldsymbol\Sigma \in \mathbb{C}^{N\times N}$, which in general requires $\mathcal{O}(N^2)$ storage and arithmetic;
\item Extraction of the covariance sub-block $\boldsymbol\Sigma_{\mathcal O\mathcal O}$, followed by inversion of an $M\times M$ matrix, whose computational complexity is estimated at $\mathcal{O}(M^3)$.
\end{itemize}

\vspace{-2mm}
\subsection{Lower Bound of Minimum Required Observations}\label{subsec:fundamental_bound}
While empirical sampling strategies provide feasible solutions for channel reconstruction, a fundamental theoretical question remains: \emph{What is the absolute minimum number of observations $M$ required to achieve a target estimation error $\varepsilon$?} We address this from an information-theoretic perspective by analyzing the spatial degrees of freedom (DoF) intrinsic to the channel covariance matrix $\boldsymbol{\Sigma}$. We establish a rigorous necessary condition (a theoretical lower bound) by evaluating the optimal performance limits of the MMSE estimator.

Recall that the channel vector $\boldsymbol{g} \in \mathbb{C}^{N}$ is modeled as a zero-mean circularly-symmetric complex Gaussian random vector with covariance matrix $\boldsymbol{\Sigma}$. According to the Karhunen-Lo\'{e}ve Theorem (KLT), $\boldsymbol{\Sigma}$ admits the eigenvalue decomposition (see (\ref{eq:channel_eigenvalue}) in Section \ref{sec:ARp_approx}), where $\boldsymbol{U} = [\boldsymbol{u}_1, \boldsymbol{u}_2, \dots, \boldsymbol{u}_N] \in \mathbb{C}^{N \times N}$ denotes the unitary matrix of eigenvectors, and $\boldsymbol{\Lambda} = \mathrm{diag}(\lambda_1, \lambda_2, \dots, \lambda_N)$ contains the eigenvalues. Without loss of generality, we assume the eigenvalues are sorted in descending order, i.e., $\lambda_1 \ge \lambda_2 \ge \dots \ge \lambda_N \ge 0$. The total energy (variance) of the channel is given by the trace of $\boldsymbol{\Sigma}$, i.e.,
\begin{equation}
\mathrm{tr}(\boldsymbol{\Sigma}) = \sum_{i=1}^{N} \lambda_i.
\end{equation}

To find the minimum estimation error from $M$ measurements, we consider the idealized scenario of an \emph{unconstrained} linear measurement. Suppose we can project the channel $\boldsymbol{g}$ onto any arbitrary $M$-dimensional subspace using a measurement matrix $\boldsymbol{W} \in \mathbb{C}^{M \times N}$, yielding observations $\boldsymbol{y}_{\mathrm{ideal}} = \boldsymbol{W}\boldsymbol{g}$. According to low-rank matrix approximation theory, the optimal choice for $\boldsymbol{W}$ that minimizes the reconstruction mean squared error is to select the eigenvectors corresponding to the $M$ largest eigenvalues, i.e., $\boldsymbol{W}_{\mathrm{opt}} = [\boldsymbol{u}_1, \dots, \boldsymbol{u}_M]^{\mathsf{H}}$. 

Notably, we need to emphasize that $\boldsymbol{W}$ is different to the binary selection matrix $\mathbf S_{\mathcal O}$ in \eqref{eq.selection_matrix}. To clarify the physical meaning of $\boldsymbol{W}$, consider a toy example with $N=4$ available antenna ports and an allowance of $M=2$ measurements. If we use the physical port selection matrix $\mathbf{S}_{\mathcal{O}}$, we are strictly operating like physical switches. For instance, selecting ports $1$ and $3$ yields the matrix $
\mathbf{S}_{\mathcal{O}} = 
\begin{bmatrix} 
1 & 0 & 0 & 0 \\ 
0 & 0 & 1 & 0 
\end{bmatrix}$,
which produces the observation $\boldsymbol{y} = [g_1, g_3]^{\mathsf{T}}$. The signals impinging on ports $2$ and $4$ are completely discarded. In contrast, the ideal unconstrained measurement matrix $\boldsymbol{W} \in \mathbb{C}^{2 \times 4}$ is assumed to be a perfect analog/digital combiner capable of using arbitrary complex weights. Suppose the strongest spatial DoF (the principal eigenvector $\boldsymbol{u}_1$) is uniformly distributed across the array, e.g., $\boldsymbol{u}_1 = [0.5, 0.5, 0.5, 0.5]^{\mathsf{T}}$. By setting the first row of $\boldsymbol{W}$ to $\boldsymbol{u}_1^{\mathsf{H}}$, the first ideal observation becomes a perfectly matched weighted sum: $y_{\mathrm{ideal},1} = 0.5g_1 + 0.5g_2 + 0.5g_3 + 0.5g_4$. In this way, $\boldsymbol{W}$ harvests energy and information from the \emph{entire} array simultaneously into $M$ outputs. This perfectly aligned combination achieves the best reconstruction performance, which is technically different from the on/off selection $\mathbf{S}_{\mathcal{O}}$.

In this unconstrained optimal case, the $M$ largest principal components are perfectly recovered, and the residual reconstruction error stems entirely from the discarded $N-M$ spatial components. Thus, the minimum achievable NMSE for \emph{any} rank-$M$ unconstrained linear measurement is exactly the normalized sum of the truncated eigenvalue tail:
\begin{equation}\label{eq:nmse_ideal_pca}
\mathrm{NMSE}_{\mathrm{ideal}}(M) = \frac{\sum_{i=M+1}^{N} \lambda_i}{\sum_{i=1}^{N} \lambda_i}.
\end{equation}

Now, we return to our practical FAS setup. The port selection operation is mathematically modeled by the selection matrix $\mathbf{S}_{\mathcal{O}} \in \{0,1\}^{M \times N}$, which restricts the measurement to $M$ physical discrete spatial points. This is fundamentally a \emph{constrained} linear measurement (often mathematically referred to as the Column Subset Selection Problem). Because constrained optimization cannot outperform unconstrained optimization, the reconstruction error of the MMSE estimator under \emph{any} specific port selection $\mathcal{O}$ with cardinality $|\mathcal{O}| = M$ must be strictly bounded below by the ideal unconstrained error. As a result, we arrive at a rigorous inequality:
\begin{equation}\label{eq:nmse_bound_inequality}
\mathrm{NMSE}(\mathcal{O}) \ge \mathrm{NMSE}_{\mathrm{ideal}}(M) = \frac{\sum_{i=M+1}^{N} \lambda_i}{\mathrm{tr}(\boldsymbol{\Sigma})},
\end{equation}
which reveals that the estimation error is fundamentally bottlenecked by the intrinsic spatial DoF of the channel. To guarantee that the target error threshold is met (i.e., $\mathrm{NMSE} \le \varepsilon$), the truncated eigenvalue tail must not exceed $\varepsilon \mathrm{tr}(\boldsymbol{\Sigma})$. Hence, by isolating $M$ from this necessary condition, we obtain the minimum required number of observations:
\begin{equation}\label{eq:M_lower_bound_final}
M_{\min}^{\mathrm{MMSE}}(\varepsilon) = \min \left\{ M \in \mathbb{N} \;\Bigg|\; \frac{\sum_{i=M+1}^{N} \lambda_i}{\mathrm{tr}(\boldsymbol{\Sigma})} \le \varepsilon \right\}.
\end{equation}

This fundamental lower bound provides an intuitive physical insight for FAS dimensioning. In highly correlated channels, such as those modeled by Clarke's correlation \eqref{eq.Clarke}, the actual useful information is highly concentrated. Mathematically, this means only the first few eigenvalues are significantly large, while the remaining eigenvalues drop rapidly to near zero, which corresponds exactly to the findings in \cite{FAS_Survey}.

The number of these dominant eigenvalues represents the effective spatial DoF of the channel. Once the number of observed ports $M$ covers this effective DoF, the residual error (which is the sum of the remaining near-zero eigenvalues) becomes vanishingly small. Because of this, even if we demand an extremely stringent estimation accuracy (i.e., $\varepsilon \to 0$), the required number of observations $M$ will not explode to infinity. Instead, it scales gracefully and plateaus near the channel's effective DoF. This explains why an FAS does not need an excessively large number of active RF chains to achieve near-perfect channel reconstruction, offering a solid and intuitive mathematical foundation for hardware design.

\vspace{-2mm}
\subsection{Low-Complexity Globally Optimal Interpolation}\label{subsec:kalman_completion}
Under the linear-Gaussian AR($p$) model introduced earlier, channel interpolation can be performed via Kalman filtering and smoothing. For linear state-space models driven by Gaussian noise, Kalman filtering yields the optimal MMSE estimator \cite{Grewal2008,Kovvali2013}. Thus, when the channel is modeled by the fitted AR($p$) process, Kalman smoothing provides a computationally efficient method that achieves the same globally optimal solution
as direct joint Gaussian conditioning.

{\em State-Space Representation via AR($p$)}---We define the $p$-dimensional state vector $\boldsymbol s_k$ in (\ref{eq:state_def}). Then the AR($p$) recursion (\ref{eq:ARp_def}) can be written as a first-order linear dynamical system
\begin{equation}\label{eq:ssm_state}
\boldsymbol s_{k+1}=\boldsymbol A \boldsymbol s_k+\boldsymbol{\varepsilon}_{k+1},	
\end{equation}
where
\begin{equation}\label{eq:companion matrix}
	\boldsymbol A =
	\begin{bmatrix}
		\alpha_1 & \alpha_2 & \cdots & \alpha_p \\
		1 & 0 & \cdots & 0 \\
		0 & 1 & \cdots & 0 \\
		\vdots & & \ddots & \vdots \\
		0 & \cdots & 1 & 0
	\end{bmatrix},
\end{equation}
is the companion matrix $\boldsymbol A \in \mathbb{C}^{p\times p}$ and
\begin{equation}
\boldsymbol{\varepsilon}_k =\begin{bmatrix}
\varepsilon_k \\
0 \\
\vdots \\
0
\end{bmatrix},~
\varepsilon_k \sim \mathcal{CN}(0,\sigma_\varepsilon^2).
\end{equation}
is the process noise with covariance given by
\begin{equation}\label{eq:process-noise covariance}
\boldsymbol Q\triangleq \mathbb{E}[\boldsymbol{\varepsilon}_k \boldsymbol{\varepsilon}_k^{\mathsf H}]=
\begin{bmatrix}
\sigma_\varepsilon^2 & 0 & \cdots & 0 \\
0 & 0 & \cdots & 0 \\
\vdots & & \ddots & \vdots \\
0 & 0 & \cdots & 0
\end{bmatrix}.
\end{equation}

{\em Arbitrary-Sparse Ports Observation Model}---At port index $k$, an observation may or may not be available. When observed, the measurement equation is given as
\begin{equation}\label{eq:ssm_obs}
y_k=\boldsymbol H \boldsymbol s_k+v_k,
\end{equation}
which follows from the model in (\ref{eq:noisy_observe}) earlier. But here, $\boldsymbol H = [1,\,0,\,\dots,\,0]$ is the selection vector that picks $g_k$ from the state vector $\boldsymbol s_k$. For noise-free observations, one may set $\sigma_v^2$ to a very small positive value for numerical stability. If $k \notin \mathcal O$, then the measurement update is skipped.

{\em Kalman Filtering (Forward Pass)}---Let $\boldsymbol s_{k|k}$ and $\boldsymbol P_{k|k}$ denote the filtered posterior mean and covariance, conditioned on observations up to time $k$. Perform the following steps:
\begin{enumerate}
\item {\bf Prediction step}---perform
\begin{subequations}
\begin{align}
	\boldsymbol s_{k|k-1}
	&=
	\boldsymbol A \boldsymbol s_{k-1|k-1}, \\
	\boldsymbol P_{k|k-1}
	&=
	\boldsymbol A \boldsymbol P_{k-1|k-1} \boldsymbol A^{\mathsf H}
	+
	\boldsymbol Q.
\end{align}
\end{subequations}
\item {\bf Update step}---If $k\in\mathcal O$,
\begin{subequations}
\begin{align}
	\boldsymbol S_k
	&=
	\boldsymbol H \boldsymbol P_{k|k-1} \boldsymbol H^{\mathsf H}
	+
	\sigma_v^2, \\
	\boldsymbol K_k
	&=
	\boldsymbol P_{k|k-1}
	\boldsymbol H^{\mathsf H}
	\boldsymbol S_k^{-1}, \\
	\boldsymbol s_{k|k}
	&=
	\boldsymbol s_{k|k-1}
	+
	\boldsymbol K_k
	\big(
	y_k - \boldsymbol H \boldsymbol s_{k|k-1}
	\big), \\
	\boldsymbol P_{k|k}
	&=
	(\boldsymbol I - \boldsymbol K_k \boldsymbol H)
	\boldsymbol P_{k|k-1}.
\end{align}
\end{subequations}
If $k \notin \mathcal O$, then $\boldsymbol s_{k|k} = \boldsymbol s_{k|k-1}$ and $\boldsymbol P_{k|k} = \boldsymbol P_{k|k-1}$.
\end{enumerate}

{\em Rauch-Tung-Striebel (RTS) Smoothing (Backward Pass)}---Filtering only incorporates past observations. To obtain the globally optimal estimate conditioned on all measurements $\{g_k\}_{k\in\mathcal O}$, we resort to RTS smoothing \cite{RTS1,RTS2}. Define $\boldsymbol s_{k|N}$ and $\boldsymbol P_{k|N}$ as the smoothed posterior mean and covariance conditioned on all the observations over $\{1,2,\ldots, N\}$, which performs the following steps:
\begin{enumerate}
\item {\bf Initialization}---At $k=N$, set
\begin{equation}
\left\{\begin{aligned}
\boldsymbol s_{N|N} &= \boldsymbol s_{N|N},\\
\boldsymbol P_{N|N} &= \boldsymbol P_{N|N}.
\end{aligned}\right.
\end{equation}

\item {\bf Recursive computation}---For $k=N-1,\dots,1$,
\begin{subequations}
\begin{align}
	\boldsymbol G_k
	&=
	\boldsymbol P_{k|k}
	\boldsymbol A^{\mathsf H}
	(\boldsymbol P_{k+1|k})^{-1}, \\
	\boldsymbol s_{k|N}
	&=
	\boldsymbol s_{k|k}
	+
	\boldsymbol G_k
	\big(
	\boldsymbol s_{k+1|N}
	-
	\boldsymbol s_{k+1|k}
	\big), \\
	\boldsymbol P_{k|N}
	&=
	\boldsymbol P_{k|k}
	+
	\boldsymbol G_k
	\big(
	\boldsymbol P_{k+1|N}
	-
	\boldsymbol P_{k+1|k}
	\big)
	\boldsymbol G_k^{\mathsf H}.
\end{align}
\end{subequations}
\end{enumerate}

{\em Recovered Channel and Uncertainty}---The reconstructed MMSE channel coefficient at port $k$ is given by
\begin{equation}\label{eq:completion_est}
\hat g_k=\mathbb{E}[g_k | \{g_m\}_{m\in\mathcal O}]=\boldsymbol H \boldsymbol s_{k|N},
\end{equation}
and the corresponding posterior variance is
\begin{equation}\label{eq:completion_var}
\mathrm{Var}(g_k| \{g_m\}_{m\in\mathcal O})=\boldsymbol H \boldsymbol P_{k|N} \boldsymbol H^{\mathsf H}.
\end{equation}
As a result, the smoother provides both point estimates and calibrated uncertainty quantification for each port.

{\em Equivalence and Complexity}---Under linear-Gaussian assumptions, Kalman smoothing yields exactly the same conditional mean and covariance as the direct Gaussian conditioning formulas, \eqref{eq:cond_mean_clean} and \eqref{eq:cond_cov_clean}. The difference lies in the computational strategy. All filtering and smoothing operations involve $p\times p$ matrices. Since $p$ is typically much smaller than $N$, the overall computational complexity scales as $\mathcal{O}(N p^2)$, which is in linear order to $N$. This represents a substantial reduction compared to the covariance-based implementation, whose cost scales at least quadratically and typically cubically to $N$.

The filter requires an initial distribution for $\boldsymbol s_1$. A natural choice is the stationary distribution implied by the AR($p$) model, i.e., $\boldsymbol s_1\sim\mathcal{CN}(\boldsymbol 0,\boldsymbol P_\infty)$, where $\boldsymbol P_\infty$ is the solution to the discrete Lyapunov equation \cite{Lyapunov equation}
\begin{equation}\label{eq:Lyapunov}
\boldsymbol {P}_\infty=\boldsymbol {A}\boldsymbol {P}_\infty\boldsymbol {A}^{\mathsf H}+\boldsymbol {Q}.
\end{equation}

This initialization ensures consistency with the fitted AR($p$) model and preserves stationarity. Since the fitted AR($p$) model is Schur-stable, (\ref{eq:Lyapunov}) admits a unique positive semidefinite solution, which equals the stationary state covariance and is therefore used to initialize the Kalman filter.

\begin{table}[]
\centering
\caption{Comparison of Port Selection Strategies}\label{tab:selections}
	\renewcommand{\arraystretch}{1.1}
	\setlength{\tabcolsep}{4pt}
\resizebox{.9\linewidth}{!}{
	\begin{tabular}{lccc}
		\hline
		Strategy & $L_{\max}$ & Extrapolation to Ends & Scaling \\
		\hline
		Random 
		& $\sim \frac{N}{M}\log M$ 
		& Possible 
		& Suboptimal \\
		
		Uniform (Endpoints) 
		& $\frac{N-1}{M-1}$ 
		& No 
		& Optimal \\
		
		Uniform (No End) 
		& $\ge \frac{N-1}{M-1}$ 
		& Possible 
		& Intermediate \\
		\hline
	\end{tabular}}
\vspace{-3mm}
\end{table}

\vspace{-2mm}
\subsection{Exemplified Port Selection Strategies}\label{subsec:selection_strategies}
Consider two typical port selection strategies (random selection and uniform selection). For an arbitrary observation set $\mathcal{O}=\{k_1<k_2<\dots<k_M\}$ drawn from $N$ ports, let $L_{\max}$ denote the maximum inter-sample gap length (i.e., the maximum missing block size). The geometrical properties of these strategies fundamentally affect the interpolation performance:

{\em Random Selection}---The observation set is drawn uniformly and randomly from all $\binom{N}{M}$ subsets. The inter-sample gaps are random variables with mean spacing approximately $N/M$. For random selection, selecting $M$ observation indices uniformly over $\{1,\dots,N\}$ is equivalent (after normalization by $N$) to drawing $M$ independent samples $U_1,\dots,U_M \sim \mathrm{Unif}(0,1)$. Let $U_{(1)}\le\dots\le U_{(M)}$ denote the order statistics and define the spacings as $S_i = U_{(i)} - U_{(i-1)}$, for $i=1,\dots,M+1,$ with $U_{(0)}=0$ and $U_{(M+1)}=1$. The vector $(S_1,\dots,S_{M+1})$ follows a Dirichlet$(1,\dots,1)$ distribution, and the largest gap corresponds to the extreme spacing $S_{\max} = \max_{1\le i\le M+1} S_i$ where as $M\to\infty$, $\mathbb{E}[S_{\max}] \sim \frac{\log M}{M}$. Since the discrete problem on $\{1,\dots,N\}$ is obtained by scaling the unit interval by $N$, the expected largest gap length satisfies
\begin{equation}
\mathbb{E}\bigl[L_{\max}\bigr] \sim \frac{N}{M}\log M.
\end{equation}
This logarithmic growth implies that random selection frequently leaves large unobserved continuous blocks, inevitably leading to suboptimal worst-case reconstruction errors.

{\em Uniform Selection}---The $M$ observed ports are placed approximately equispaced over $\{1,\dots,N\}$ to minimize the maximum inter-sample gap. Let the average spacing be $\Delta \approx \frac{N}{M}$. A general equispaced construction can be written as
\begin{equation}
k_i \approx a + (i-1)\Delta,~i=1,\dots,M,
\end{equation}
in which the offset $a$ plays the role of deciding whether the boundary ports (i.e., endpoints) are included.

{\bf Case (i): Endpoints included}---When $a=1$, it is enforced that $k_1=1$, $k_M=N$, yielding $\Delta = \frac{N-1}{M-1}$, and hence
\begin{equation}
k_i \approx 1+(i-1)\frac{N-1}{M-1}.
\end{equation}
In this case, all the gaps, including those from the endpoints, are equalized, and $L_{\max} \le \left\lceil\frac{N-1}{M-1}\right\rceil$. \emph{No boundary extrapolation} is required, as the extreme ports at two ends are directly observed. This configuration minimizes the maximum gap in a minimax sense (e.g., for $N=10$, $M=4$, $\mathcal{O}=\{1,4,7,10\}$), offering the widest placement for spatial interpolation.

{\bf Case (ii): Endpoints not enforced}---If the grid is shifted inward (e.g., $a>1$), the interior spacing remains approximately $\Delta$, but additional boundary gaps of size $k_1-1$ and $N-k_M$ appear. These boundary gaps require extrapolation rather than interpolation. Consequently, the maximum missing block length becomes
\begin{equation}
L_{\max} = \max\{\text{interior gaps},\; k_1-1,\; N-k_M\},
\end{equation}
which may exceed $\lceil (N-1)/(M-1) \rceil$. Performance degradation stems from these enlarged boundary gaps requiring extrapolation (e.g., for $N=10$, $M=4$, $\mathcal{O}=\{3,5,7,9\}$).

To clarify, we summarize the geometrical properties of the exemplified port selection strategies in Table~\ref{tab:selections}. \emph{However, we encourage our readers to explore optimal port selection design in future works, as the interpolation NMSE can be rendered directly as a function of the sampling matrix.}

\begin{figure}[]
\centering
\includegraphics[width=\columnwidth]{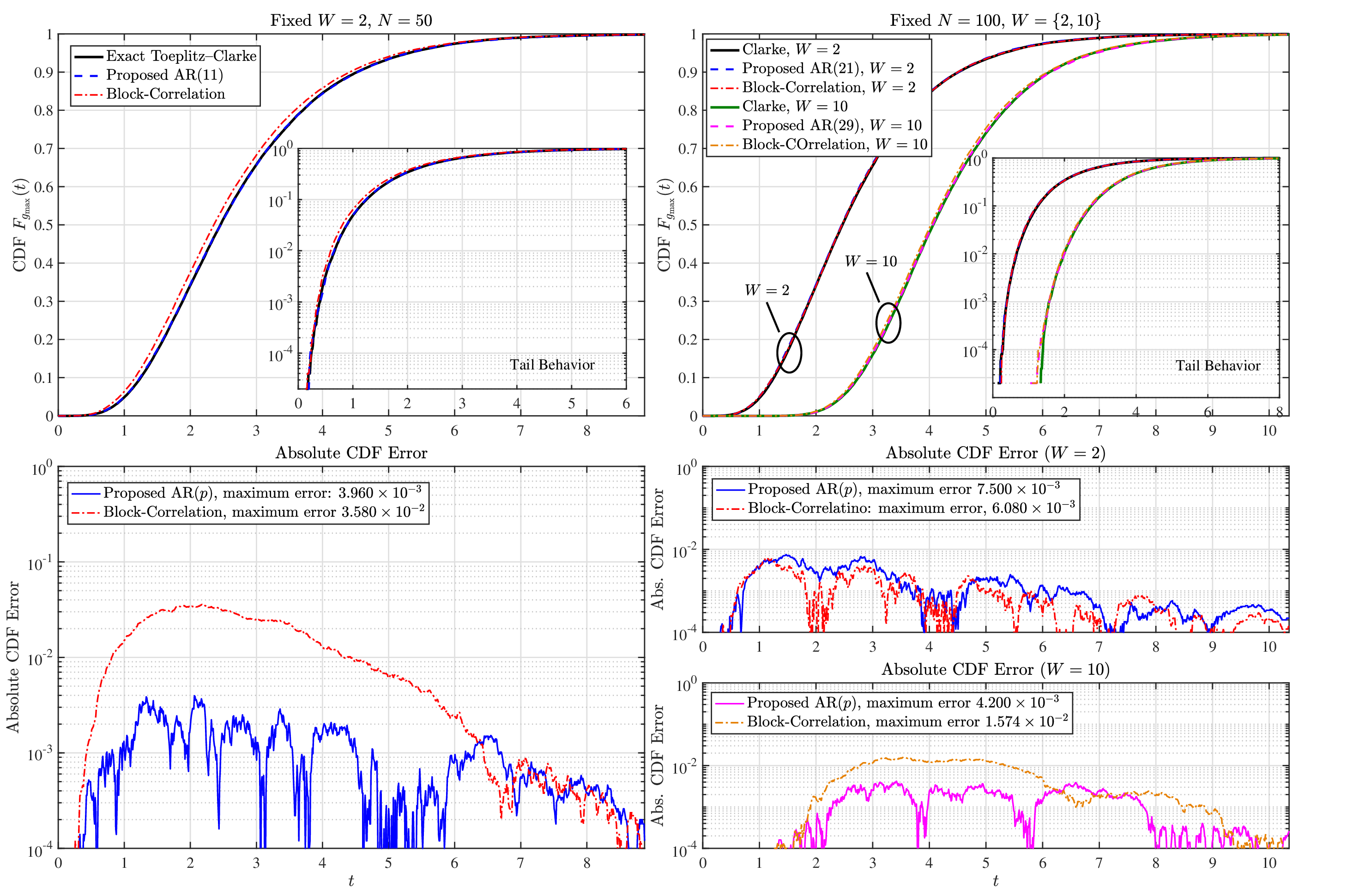}
\caption{Correlation modeling comparison between the AR($p$) framework and block-correlation approximation \cite{Clarke3} under different number of ports $N\in\{50, 100\}$ and different FAS size $W\in \{2,10\}$. For block-correlation approximation, the eigenvalue threshold is set as $1$. For the proposed AR($p$) framework, the maximum tolerable approximation order is fixed to $p^{\star}=40$. All CDFs are estimated based on $5\times 10^4$ Monte-Carlo samples.}\label{fig:Error_VS_Block_Correlation}
\vspace{-4mm}
\end{figure}

\begin{figure*}[]
\centering
\includegraphics[width=\linewidth]{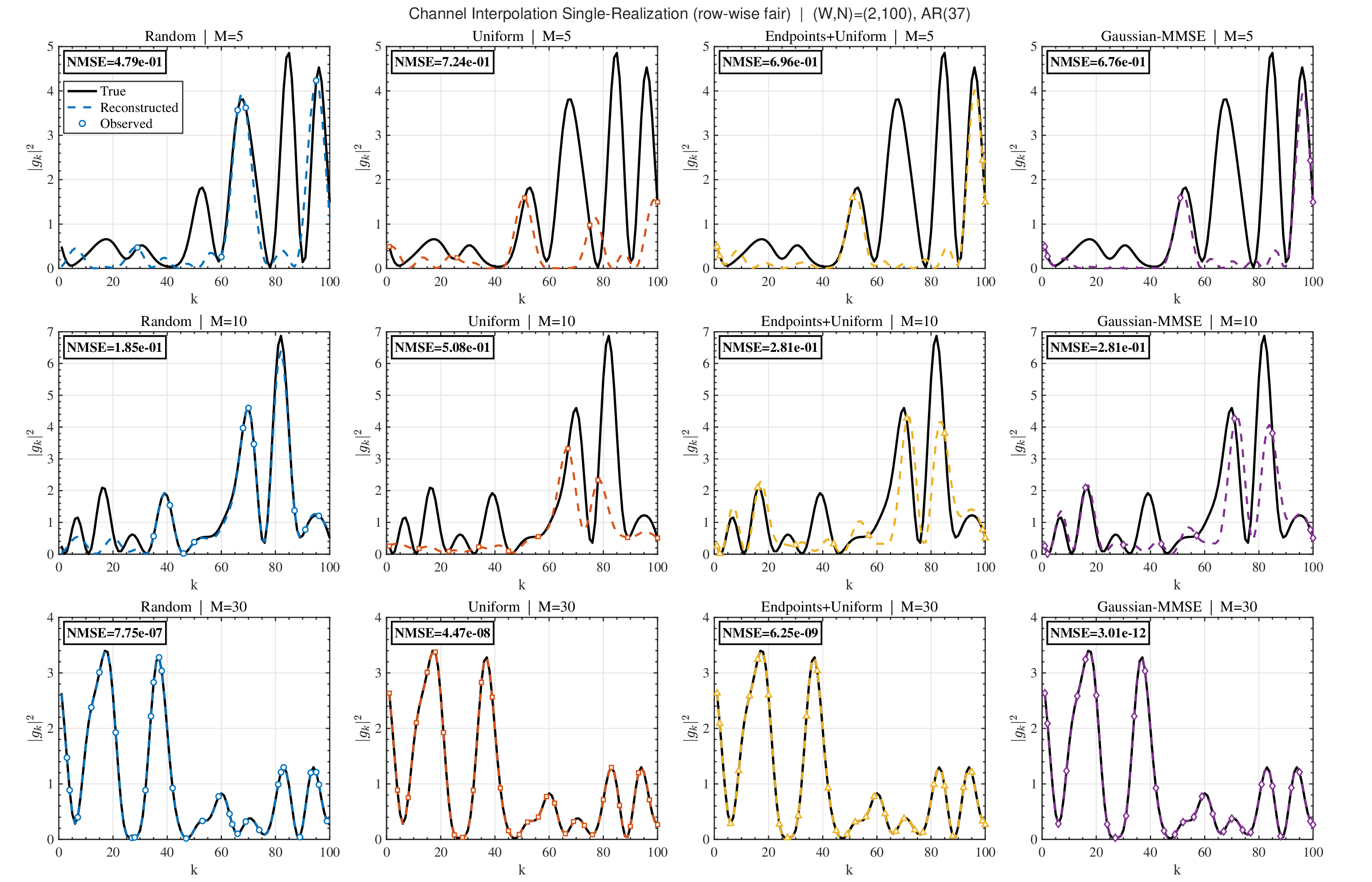}
\caption{One realization of channel interpolation with different port selection strategies under $(W,N)=(2,100)$ and $M\in\{5,10,30\}$.}\label{fig:single_realization_4_selection_strategies}
\vspace{-4mm}
\end{figure*}

\begin{figure}[]
\centering
\includegraphics[width=0.9\columnwidth]{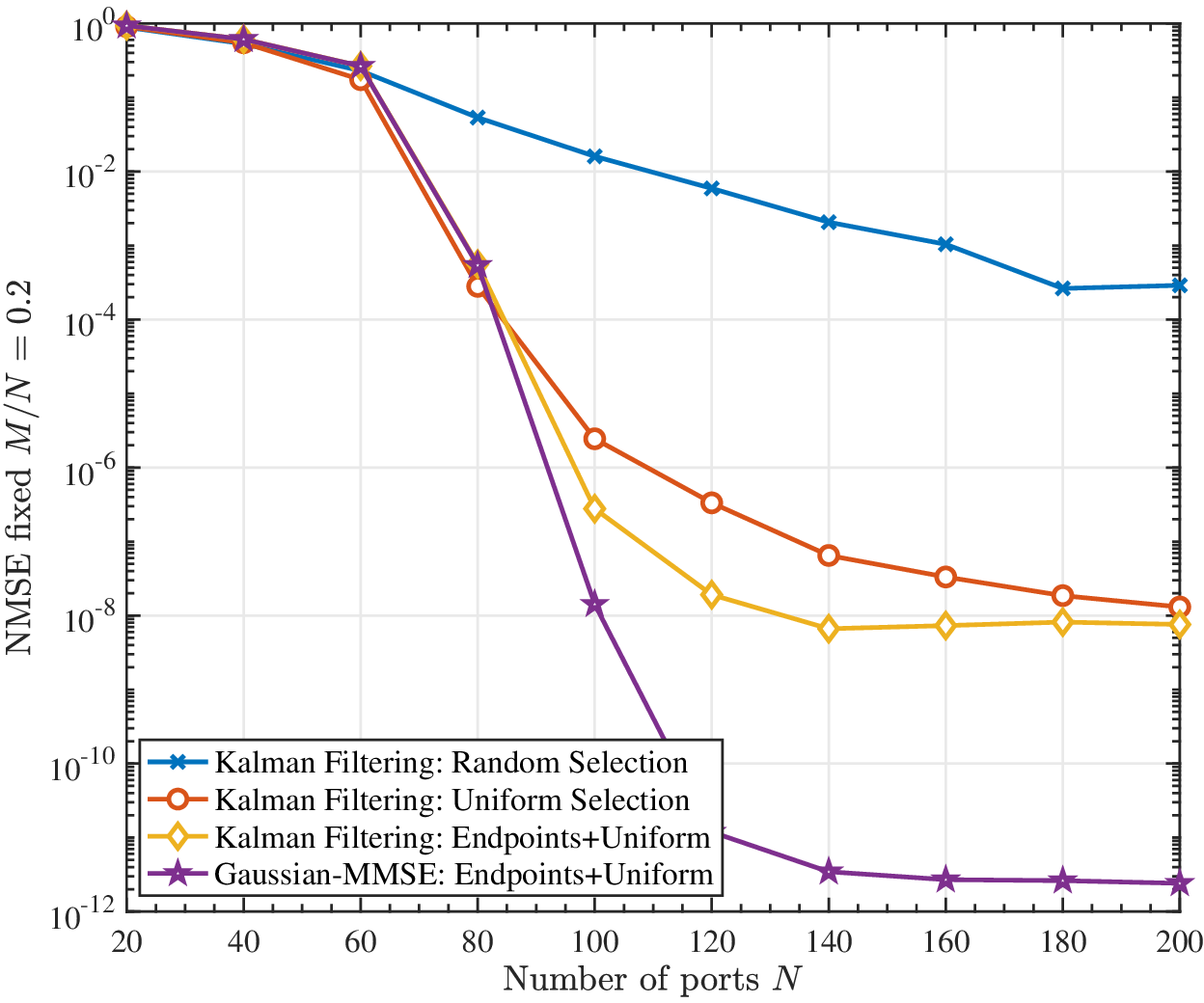}
\caption{Illustration of channel interpolation NMSE under fixed ratio of observation/total port ($M/N=0.2$) under $W=2$, AR($p$) order upper-bound $p_{\max}=40$. To the Gaussian-MMSE estimator, the exact correlation matrix is a known prior. To the Kalman Filtering estimator, the interpolation is conducted under approximated correlation modeling via AR($p$).}\label{fig:NMSE_fixed_Observations_Ratio}
\vspace{-4mm}
\end{figure}

\begin{figure}[]
\centering
\includegraphics[width=0.95\columnwidth]{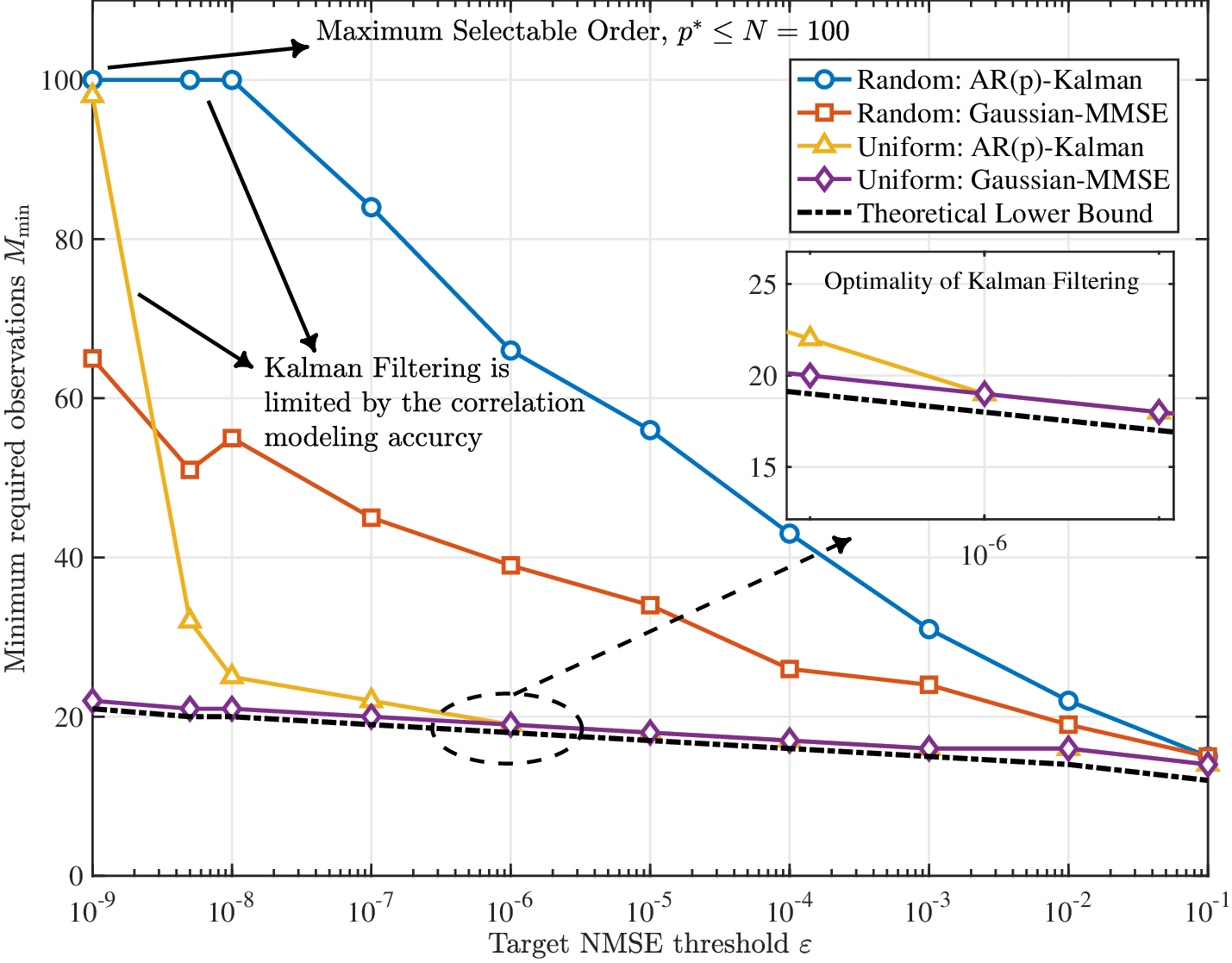}
\caption{Illustration of empirical and theoretical results for minimum-required observation count to reach a target NMSE $\varepsilon$ under $(W,N)=(2,100)$ with tolerable maximum order $p^{\star}=40$. For Gaussian-MMSE, the correlation matrix of Clarke's modeling is fully known, while for Kalman Filtering, the correlation modeling is fitted via AR($p$) with adaptive order selection.}\label{fig:Minimum_required_Observations}
\vspace{-4mm}
\end{figure}

\vspace{-2mm}
\section{Numerical Results}\label{sec.numerical}
In this section, numerical results are presented to validate the proposed correlation modeling framework. Specifically, we compare the accuracy of correlation approximation between the proposed AR($p$) method and the block-correlation method \cite{Clarke3,Clarke4}. After that, we compare the performance of Kalman filtering under the AR($p$) modeling with the Gaussian-MMSE estimator with fully known prior of the Clarke's correlation matrix depicting the best interpolation performance. 

{\em Correlation Modeling Comparison}---Fig.~\ref{fig:single_realization_4_selection_strategies} demonstrates the correlation modeling comparison between the proposed AR($p$) framework and block-correlation approximation \cite{Clarke3} under different port count $N\in\{50, 100\}$ and different FAS size $W\in \{2,10\}$. The results show as expected that block-correlation modeling relies on the sparsity in the eigenvalue space which is only accurate when $N$ is relatively large. At $(W,N)=(2,50)$, the maximum approximation error of block-correlation is $10$ times larger than the proposed AR($p$) model while achieving similar performance to the proposed method at $(W,N)=(2,100)$. At $(W,N)=(5,100)$, the accuracy of the block-correlation deteriorates again. 

{\em CSI Interpolation Comparison}---Fig.~\ref{fig:single_realization_4_selection_strategies} shows one particular realization of channel interpolation using different port selection strategies under $(W,N)=(2,100)$ and $M\in\{5,10,30\}$. Despite the complex fading structures, all considered port selection strategies can interpolate the channel reliably. Random selection achieves the worst NMSE while `Endpoints + Uniform with Gaussian MMSE' performs the best. Furthermore, Fig.~\ref{fig:NMSE_fixed_Observations_Ratio} compares the NMSE of different port selection strategies with a fixed observation ratio $\frac{M}{N}=0.2$. All curves eventually converge to an NMSE floor indicating that the performance is ultimately limited by the fixed ratio $\frac{M}{N}=0.2$.

{\em Observation Count Validation}---Fig.~\ref{fig:Minimum_required_Observations} illustrates the predicted lower-bound of the minimum required number of observations $M_{\min}$ against the NMSE target $\varepsilon$ when $(W,N)=(2,100)$. The results validate the proposed lower bound on $M_{\min}$ since the theoretical lower-bound is tight to the empirical curve of MMSE. Additionally, Kalman filtering driven by the AR($p$) modeling is proved to be the optimum estimator, as long as the target NMSE error is no smaller than the precision of correlation approximation via the AR($p$) modeling.

\vspace{-2mm}
\section{Conclusion}\label{sec.conclusion}
This paper developed a unified framework for correlation approximation and channel interpolation in FASs. By introducing a $p$-order AR generative model, we replaced conventional covariance-based representations with a flexible, complexity-scalable alternative that is capable of modeling any correlation structures. The proposed model enables controllable accuracy-complexity tradeoffs, establishes the fundamental limits of CSI interpolation under arbitrary observations, and presents efficient algorithms for channel reconstruction.

\end{document}